\documentclass[fleqn,usenatbib]{mnras}

\usepackage{newtxtext,newtxmath}

\usepackage[T1]{fontenc}
\usepackage{ae,aecompl}

\usepackage{graphicx}	
\usepackage{amsmath}	
\usepackage{amssymb}	
\usepackage{bm}
\usepackage[usenames]{color}

\graphicspath{{Figures/}}
\usepackage[normalem]{ulem}
\usepackage{array}

\newcolumntype{P}[1]{>{\centering\arraybackslash}p{#1}}
\newcolumntype{M}[1]{>{\centering\arraybackslash}m{#1}}

\title[Wave effects in microlensing]{Wave effects in the microlensing of pulsars and FRBs by point masses}

\author[Jow et al.]{
Dylan L. Jow,$^{1,2}$\thanks{E-mail: djow@physics.utoronto.ca}
Simon Foreman,$^{3,4}$
Ue-Li Pen$^{1,2,3,5,6,7}$
and Wei Zhu$^{1}$
\\
$^{1}$Canadian Institute for Theoretical Astrophysics, University of Toronto, 60 St. George Street, Toronto, ON M5S 3H8, Canada\\
$^{2}$Department of Physics, University of Toronto, 60 St. George Street, Toronto, ON M5S 1A7, Canada\\
$^{3}$Perimeter Institute for Theoretical Physics, 31 Caroline St. North, Waterloo, ON, Canada N2L 2Y5\\
$^{4}$Dominion Radio Astrophysical Observatory, 
Herzberg Astronomy \& Astrophysics Research Centre,  \\
National Research Council Canada, P.O.\ Box 248, Penticton, BC V2A 6J9, Canada\\
$^{5}$Canadian Institute for Advanced Research, CIFAR program in Gravitation and Cosmology\\
$^{6}$Dunlap Institute for Astronomy \& Astrophysics, University of Toronto, AB 120-50 St. George Street, Toronto, ON M5S 3H4, Canada\\
$^{7}$Max-Planck-Institut f{\"u}r Radioastronomie, Auf dem H{\"u}gel 69, D-53121 Bonn
}

\date{Accepted XXX. Received YYY; in original form ZZZ}

\pubyear{2019}

\begin{document}
\label{firstpage}
\pagerange{\pageref{firstpage}--\pageref{lastpage}}
\maketitle

\begin{abstract}
Wave effects are often neglected in microlensing studies; however, for coherent point-like sources, such as pulsars and fast radio bursts (FRBs), wave effects will become important in their gravitational lensing. In this paper, we describe the wave optics formalism, its various limits, and the conditions for which these limits hold. Using the simple point lens as an example, we show that the frequency dependence of wave effects breaks degeneracies that are present in the usual geometric optics limit, and constructive interference results in larger magnifications further from the lens. This latter fact leads to a generic increase in cross section for microlensing events in the wave-optics regime compared to the geometric optics regime. For realistic percent-level spectral sensitivities,
this leads to a relative boost in lensing cross section of more than an order of magnitude. We apply the point-lens model to the lensing of FRBs and pulsars and find that these radio sources will be lensed in the full wave-optics regime by isolated masses in the range of $0.1-100\,M_\oplus$, which includes free-floating planets (FFPs), whose Einstein radius is smaller than the Fresnel scale. More generally, the interference pattern allows an instantaneous determination of lens masses, unlike traditional microlensing techniques which only yield a mass inference from the event
timescale.
\end{abstract}




\section{Introduction}
\label{sec:intro}

The effect of microlensing on astrophysical sources is a powerful probe of faint, but massive foreground objects. For example, microlensing of stars in the Milky Way has led to a fruitful avenue for the detection of exoplanets \citep{1991ApJ...374L..37M, 1992ApJ...396..104G, gaudi_microlensing_2012}, as well as tight constraints on the percentage of dark matter in the galaxy that can be attributed to Massive Compact Halo Objects (MACHOs) \citep{Paczynski1986,Wyrzykowski2011}. Gravitational lensing is typically studied in the geometric limit of optics, in which light rays are taken to travel along null geodesics, and the magnification of a source is determined by how lenses focus or defocus these. The semi-classical or Eikonal approximation of optics includes wave effects to first-order by allowing photons propagating along different null geodesics connecting source to observer to interfere with each other. In contrast, in the full wave-optics picture, the propagation of light is determined by a path integral picture, in which photons traverse all possible paths, with each path weighted by a phase factor related to the proper time along the path \citep[e.g.][]{nakamura}. Path integrals are highly oscillatory integrals which are numerically challenging to compute, and since in many cases of astrophysical interest the geometric limit of optics is a good approximation \citep{Schneider}, wave effects in microlensing are rarely considered.

Nevertheless, there are cases where the geometric limit is no longer an accurate description, and the full wave optics picture must be employed. The geometric limit is effectively a high energy, or high frequency, limit of the underlying wave optics, which holds for interactions between many lenses and sources of electromagnetic radiation. However, when considering the propagation of radio frequency waves through curved space-time, such as gravitational radiation, wave effects must be taken into account \citep{Takahashi:2003ix,Dai:2018enj,DiegoHannuksela2019,Diego2019}. But even when the geometric or Eikonal limits are good approximations given the source and lens parameters, both approximations yield formally infinite magnifications near caustics. An appeal to the full wave optics picture is required to ameliorate these un-physical magnifications \citep{Jaroszynski:1995cd}. 

Not only is it the case that there are astrophysically relevant situations in which the geometric limit fails, it is also true that diffractive effects can provide more information about the lensing system. Unlike geometric optics, diffractive effects are frequency dependent, in a way that depends on the physical parameters of the lensing system. \citet{Gould:1992} proposed that observations of diffractive effects in the Eikonal regime of lensed gamma ray bursts (GRBs) could be used to constrain possible dark matter candidates with masses in the range of $10^{-16}$ to $10^{-13}\,M_\odot$. Extending the result to the full wave-optics regime, \citet{Ulmer:1994ij} showed that the range of relevant masses could be increased up to  $10^{-11}\,M_\odot$. \citet{Barnacka:2012bm} used microlensing observations of GRBs to constrain primordial black holes (PBHs) as candidates for dark matter; however, following this, \citet{Katz:2018zrn} showed that for GRBs, finite source effects (i.e. effects due to the fact that realistic sources are not point-like) dominate over wave effects, making diffractive effects difficult to observe. Thus, the lack of observations of the predicted wave optics effects could not be used to constrain dark matter candidates. Similarly, while \citet{Niikura:2019zjd} proposed that wave effects in the Eikonal limit for microlensed stars in M31 could place constraints on PBH dark matter, \citet{Sugiyama:2019dgt} argue that finite source effects for stars would make placing such constraints difficult. Further, \citet{Montero-Camacho:2019jte} find that the event rate for stars to be lensed by PBHs is overestimated in the Eikonal limit. The formal infinity near caustics leads to an overestimation of the magnification for some events compared to what would actually occur, leading to an overestimated event rate. Thus, even when diffractive effects are observationally useful, finite source effects can wash out the desired signal, and misapplication of the limits that make wave-optics calculations analytically tractable can result in misleading conclusions. 

In general, it is only for coherent point-like sources that wave effects become observable. Stars only become point-like in the relevant sense for very nearby lenses, such as those in our solar system \citep{Heyl:2010dg, Heyl:2009av, Heyl:2010hm}. Two examples of astrophysical sources that are effectively coherent and point-like are pulsars and fast radio bursts (FRBs). Both pulsars and FRBs are radio sources that have physical dimensions on the sky of order $10^{-6}$ and $10^{-12}\,\mu$as.\footnote{These numbers take a pulsar's typical size and distance to be $\mathcal{O}({\rm km})$ and $\mathcal{O}({\rm kpc})$, while assuming that FRB emission regions are similar in physical size but located at cosmological distances, of order Gpc.} Thus, wave effects will generically be observable when these sources are lensed. Indeed, \citet{Katz2019} argue that observations of diffractive effects in the lensing of FRBs could tightly constrain the abundance of dark matter in the form of MACHOs.

In this work, we will describe the full wave-optics formalism in curved space-time, its various limits, and the conditions for which each regime holds, including what is meant by ``point-like". We will use the example of the wave optics of a point lens (developed by \citet{PhysRevD.34.1708, 1986ApJ...307...30D}, and elsewhere) to motivate the study of wave optics in gravitational lensing, beyond the geometric and Eikonal limits, emphasising that the inclusion of wave optical effects indeed yields additional information about the lens, breaking degeneracies that are present in the geometric regime. We will also present the novel result that wave effects will generically increase the cross section for microlensing events relative to the usual geometric optics cross section. While we choose the specific example of the point lens as a proof-of-concept for the utility of studying wave optical effects, the lessons learned may also be applicable to more complicated lenses, including external shear or binaries. Finally, we consider applications of the point-lens model to the lensing of pulsars and FRBs, showing that they may be used to constrain populations of free-floating planets (FFPs). In the coming years, with telescopes such as the Canadian Hydrogen Instensity Mapping Experiment \citep[CHIME;][]{Amiri:2018qsq,Ng:2017djg}, the Hydrogen Intensity and Real-time Analysis eXperiment \citep[HIRAX;][]{HIRAX}, and the Square Kilometer Array \citep{Keane:2014vja,Macquart:2015uea}, in addition to possible next-generation radio telescopes such as the Canadian Hydrogen Observatory and Radio-transient Detector \citep[CHORD;][]{2019arXiv191101777V} and the Packed Ultra-wideband Mapping Array \citep[PUMA;][]{Ansari:2018ury, PUMA}, we will see a wealth of new observations of radio sources on the sky. Understanding wave effects in gravitational lensing will be an important step in turning these sources into powerful probes of the universe.

\section{Wave optics in gravitational lensing}
\label{sec:wave_optics}

\subsection{Light propagation in perturbed space-time}

\begin{figure}
    \centering
    \includegraphics[width=\columnwidth, trim = 0 40 120 50]{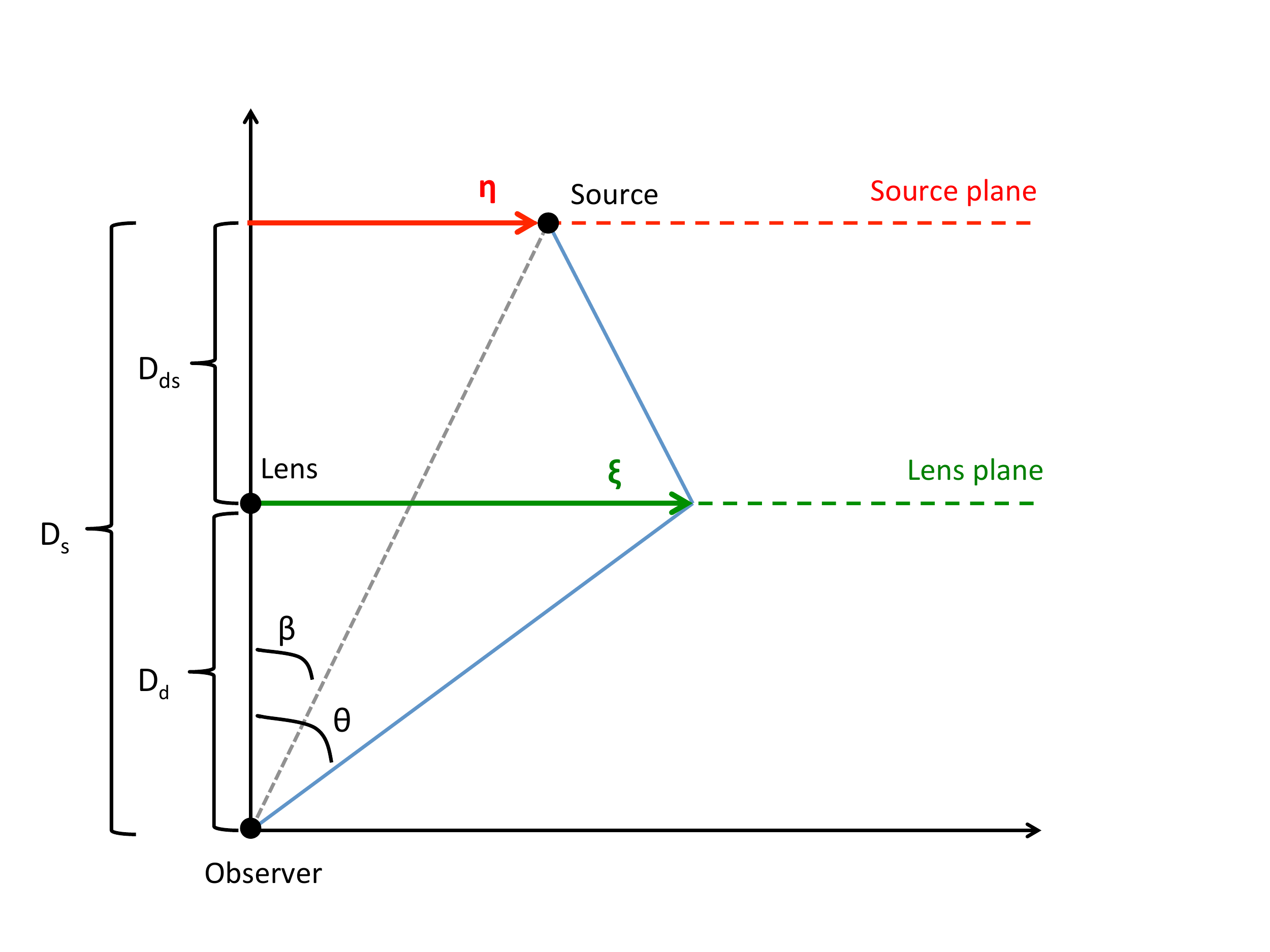}
    \caption{
    Geometry of a source at a distance $D_s$ from an observer being gravitationally lensed by a lens at a distance $D_d$ from the observer. The unperturbed line-of-sight from observer to source is shown as a gray dashed line. The vectors $\bm{\eta}$ and $\bm{\xi}$ are vectors in the source and lens plane, respectively (planes perpendicular to the optical axis, containing the source and lens). Together, $\bm{\eta}$ and $\bm{\xi}$ define a path, shown in blue, from source to observer. The radiation seen by the observer is determined by a path integral over all such paths.  Note that we consider the limit where all angles are small.
    }
    \label{fig:lens_setup}
\end{figure}

Gravitational lensing occurs when radiation propagates in a perturbed space-time. Consider a perturbed Minkowski metric, 
\begin{equation}
    ds^2  = (-1+h_{00})dt^2 + h_{0i}(dtdx^i + dx^idt) + (1+h_{ij})dx^i dx^j,
\end{equation}
where the perturbation is assumed to be small (i.e. $|h_{\mu\nu}| \ll 1$), and we have taken $c=1$. Let $S$ be some source of electromagnetic radiation, let $O$ be an observer, and let $D_s$ be the distance from the observer to the source along the optical axis. The direction and origin of the optical axis can be chosen arbitrarily, but for simplicity, here we will choose for the optical axis to begin at the observer and to extend in some direction associated with the lens. If, for example, the perturbation is sourced by a set of point masses, then the optical axis may be chosen to point from the observer to the centre-of-mass of the lens. We will assume that the lens is ``thin", by which we mean that the region in which the perturbation is non-negligible is a region perpendicular to the optical axis with thickness much less than $D_s$. This allows us to define the distances $D_d$ and $D_{ds}$: the distance from the observer to the lens and from the lens to the source, respectively. This set-up is shown in Fig.~\ref{fig:lens_setup}.

To describe lensing in this setup, we will make use of the Fermat potential $T_\Gamma$. For a trajectory $\Gamma$ connecting source to observer, $T_\Gamma$ is defined as the difference in time taken by light to traverse the trajectory in the perturbed space-time versus the time taken to traverse the null geodesic in the background space-time. (One can interpret this as the difference in arrival time between two photons emitted by the source at the same time in each space-time). To compute the Fermat potential, we first compute the time taken by a photon to traverse some path $\Gamma = x^i(\lambda)$ through the perturbed space-time:
\begin{equation}
    t_\Gamma = L_\Gamma + \frac{1}{2} \int_\Gamma d\lambda (h_{00} + h_{ij} n^i n^j + 2h_{0i}n^i)
\end{equation}
where $L_\Gamma$ is the length of the path, and $n^i = dx^i/d\lambda$ is the unit vector tangent to the path. Thus, the Fermat potential is given by
\begin{equation}
    T_\Gamma = t_\Gamma - t_0 = \Delta L_\Gamma + \frac{1}{2}\int_\Gamma d\lambda (h_{00} + h_{ij} n^i n^j + 2h_{0i}n^i),
\label{eq:Tgamma}
\end{equation}
where $\Delta L_\Gamma$ is the difference in path length between $\Gamma$ and the straight line connecting source to observer. The $\Delta L_\Gamma$ term is known as the ``geometric part" of the time delay. 

The quantity $\omega T_\Gamma$ then tells us the difference in the phase of the radiation at the observer in the perturbed versus un-perturbed case. Here and elsewhere we denote the angular frequency of the light by $\omega \equiv 2\pi \nu$, where $\nu$ is the frequency. In the thin lens approximation described above, $T_\Gamma$ is only a function of $\bm{\eta}$, the vector in the source plane specifying the location of the source, and $\bm{\xi}$, a vector in the lens plane (see Fig.~\ref{fig:lens_setup}). That is, if the lens has negligible impact outside the lens plane, we only need to consider the space of paths defined piece-wise as straight lines from the source to some point in the lens plane, and then from the lens plane to the observer. 

Using a scalar wave analysis, \citet{nakamura} study monochromatic radiation with angular frequency $\omega$, specified by a phase and an amplitude, $\phi({\bm x}, t) = \tilde{\phi}({\bm x}) e^{i\omega t}$, propagating according to the wave equation $\partial_\mu(\sqrt{-g} g^{\mu\nu} \partial_\nu \phi)=0$. \citet{nakamura} compute the amplification factor (defined to be the ratio of the field at the observer in the presence of the lens to the field in the absence of the lens, $F(O) = \tilde{\phi}(O)/\tilde{\phi}_0(O)$) for a point source to be
\begin{equation}
    F \propto \int d^2\bm{\xi} \exp \left\{ i\omega T(\bm{\xi}, \bm{\eta}) \right\}\ .
\label{eq:FT}
\end{equation}
This is a path integral familiar from quantum mechanics, over a set of paths determined by the thin lens approximation.

We will also assume that the lens and the source are confined to some small region in the lens and source planes, respectively, and that the source is close to the optical axis, where ``small" and ``close" are defined relative to the distances $D_d$, $D_s$, and $D_{ds}$. This allows us to make the small angle approximation, where we assume that $\theta$ and $\beta$ (defined in Fig.~\ref{fig:lens_setup}) are much less than 1. Defining the vectors $\bm{\theta} = \bm{\xi}/D_d$ and $\bm{\beta} = \bm{\eta}/D_s$, the small angle approximation allows us to write the Fermat potential as
\begin{equation}
    T(\bm{\theta}, \bm{\beta}) = \frac{D_dD_s}{2D_{ds}} |\bm{\theta} - \bm{\beta}|^2 - \hat{\psi}(\bm{\theta}),
\end{equation}
where the first term is the geometric part of the time delay, and the second term, $\hat{\psi}$ comes from the integral over the perturbation in Eq.~\ref{eq:Tgamma}. Now, computing the constant of proportionality for $F$ defined so that $F=1$ in the absence of a lens, we obtain, as in~\citet{nakamura}
\begin{align}
\nonumber
    F(\omega, \bm{\beta}) &= \frac{\omega}{2\pi i} \frac{D_dD_s}{D_{ds}} \int d^2\bm{\theta} \exp \big\{ i\omega \big(  \frac{D_dD_s}{2D_{ds}} |\bm{\theta} - \bm{\beta}|^2 - \hat{\psi}(\bm{\theta})\big) \big\} \\
    &= \frac{1}{2\pi i \theta^2_F} \int d^2\bm{\theta} \exp\big\{ \frac{i}{2\theta^2_F} \big(|\bm{\theta} - \bm{\beta}|^2 - \frac{2}{D} \hat{\psi}(\bm{\theta}) \big) \big\},
\label{eq:F}
\end{align}
where we have introduced the quantity $D = D_dD_s/D_{ds}$, and the Fresnel scale, 
\begin{equation}
    \theta_F \equiv \sqrt{\frac{1}{\omega D}} = \sqrt{\frac{\lambda}{2\pi D}}.
    \label{eq:frnl}
\end{equation}

The result for an extended source is obtained simply by integrating over the extent of the source, i.e. $F(\omega) = \int d^2\bm{\eta} F(\omega, \bm{\eta}) \tilde{\phi}(\bm{\eta})$. Similarly, we can integrate over frequency to obtain the result for non-monochromatic waves. From this, the magnification of the source as measured by the observer is simply $H = |F|^2$.

The Fresnel scale defined in Eq.~\eqref{eq:frnl} will be fundamental in diagnosing the importance of wave optical effects in the lensing situations we discuss later in this paper. For now, we note that it has no dependence on the properties of the perturbation to the time delay $\hat{\psi}$, but only on the ratio of the incident light's wavelength $\lambda$ and the distance ratio $D$. Note also that the above result is given for a background Minkowski space-time. For cosmological situations, the result must be computed for a background FLRW space-time. Fortunately, the result is the same, except that the distances $D_d$, $D_s$, and $D_{ds}$ refer to the angular diameter distances, and the frequency $\omega$ in the exponent is replaced by $\omega (1+z_d)$, where $z_d$ is the redshift of the lens. The same result is computed by \citet{Schneider} from Maxwell's equations in curved space-time, as opposed to the simple scalar theory used by \citet{nakamura}. \citet{job_pl} obtain the same result starting with Feynman's path integral formulation of quantum mechanics. 

\subsection{Geometric and Eikonal approximations}
\label{sec:evaluation}

Equation~\ref{eq:F} gives us a formula for computing the amplitude of the electric field propagating from a distant source in the presence of an intervening gravitational lens. In principle, this is all we need to know in order to compute the effects of a general gravitational lens; however, in practice, the diffraction integral of Eq.~\ref{eq:F} is rarely computable analytically, except for in a few special cases, and is numerically challenging to integrate. The problem lies in the fact that the integral is performed over a highly-oscillatory function, resulting in a conditionally convergent integral. Numerically, such integrals are difficult to evaluate as they depend strongly on the grid spacing, and many standard numerical techniques fail when applied to them. In the field of lattice QCD simulations, this is known as the sign problem \citep{2010arXiv1005.0539D}. Attempts to ameliorate the sign problem in the context of lensing using Picard-Lefschetz theory have recently been made \citep[see e.g.][]{job_pl}; however, historically, the solution has been to use various limits in which the diffraction integral becomes more tractable. Here we will give a brief overview of two of these limits: the geometric, and the semi-classical, or Eikonal limit. 

Both the geometric and Eikonal limits utilize the fact that for rapidly oscillatory integrals of the form shown in Eq.~\ref{eq:FT}, in the limit as $\omega \to \infty$ the largest contribution to the integral comes from regions near stationary points of $T$, i.e. where $\frac{\partial T(\bm{\theta}, \bm{\beta})}{\partial \bm{\theta}}  = 0$. Let $\bm{\theta}_i$ be the set of stationary points of $T$, which we assume to be finite. Then we can expand $T(\bm{\theta})$ about the $i^\mathrm{th}$ stationary point as
\begin{equation}
    T(\bm{\theta}) = T(\bm{\theta}_i) + \frac{1}{2} \sum_{ab} \overline{\theta}_a \overline{\theta}_b \partial_a \partial_b T(\bm{\theta}_i) + \frac{1}{6} \sum_{abc} \overline{\theta}_a \overline{\theta}_b \overline{\theta}_c  \partial_a \partial_b \partial_c T(\bm{\theta}_i) + ...
    \label{eq:Texpansion}
\end{equation}
where $\bm{\overline{\theta}} \equiv \bm{\theta} - \bm{\theta}_i$, $\partial_a \equiv \partial/\partial\theta_a$, the indices in the sums run from 1 to 2, and we have suppressed the second argument of $T$ which we are keeping fixed. As $\omega \to \infty$, we can use this expansion to obtain \citep{nakamura}
\begin{equation}
    F \approx \sum_i |H(\bm{\theta}_i)|^{1/2} \exp \big[i\omega T(\bm{\theta}_i) - i\pi n_i \big],
\end{equation}
where $H(\bm{\theta}_i)$ is the magnification corresponding to each stationary point in isolation, given by
\begin{equation}
    H_i \equiv H(\bm{\theta}_i) = \frac{1}{\mathrm{det}(\partial_a \partial_b T(\bm{\theta}_i))},
\end{equation}
and $n_i = 0, 1/2, 1$ when $\bm{\theta}_i$ is a minimum, saddle point, or maximum of $T$. Reintroducing the suppressed variable $\bm{\beta}$, the total magnification $H \equiv |F|^2$ is then
\begin{align}
\begin{split}
    &H(\omega, \bm{\beta}) \approx H_{\rm Eik}(\omega, \bm{\beta}) \equiv \sum_i |H_i(\bm{\beta})|\\ 
    &+ 2 \sum_{i<j} |H_i(\bm{\beta}) H_j(\bm{\beta}) |^{1/2}
    \cos\left[ \omega \left\{ T(\bm{\theta}_j, \bm{\beta}) - T(\bm{\theta}_i, \bm{\beta}) \right\}
     - \pi(n_j - n_i) \right].
\end{split}
\label{eq:semiclassical}
\end{align}
This is the semi-classical, or Eikonal limit of the magnification, where voltages add coherently. 

We can interpret this by first identifying the paths defined by $(\bm{\theta}_i, \bm{\beta})$ with the classical paths taken by a light ray, as given by Fermat's principle of least time. That is, the paths given by the stationary points of $T$ are the null geodesics in the perturbed space-time. Each stationary point $\bm{\theta}_i$ then corresponds to an image of the source produced by the lens. We can compute the total magnification by combining the magnification of each image (given by $H_i$) and keeping track of the phase of the light along the classical paths. Then, in this limit, the total magnification is the sum of the $H_i$ plus a term representing the interference of the images with each other, since they will generally arrive at the observer with a different phase. 

This approximation works well under two conditions:
\begin{enumerate}
\item when the images are well-separated compared to the Fresnel scale, $|\bm{\theta}_i-\bm{\theta}_j| \gg \theta_F$. This follows from the stationary phase approximation applied to the integral in Eq.~\eqref{eq:F}, which tells us that contributions to the integral vanish everywhere except near stationary points of the exponent in the limit as $\theta_F \to 0$, or, in other words, when $\theta_F$ is small compared to the other angular scale in the problem, which in this case is the angular separation of the images.
\item when none of the images are near a singular point, $\mathrm{det}(\partial_a \partial_b T(\bm{\theta}_i)) = 0$. At such a point, the magnification in the Eikonal limit becomes formally infinite, requiring higher-order terms in Eq.~\eqref{eq:Texpansion} to be retained.
\end{enumerate}

The geometric optics limit ignores the second term in Eq.~\eqref{eq:semiclassical}, resulting in
\begin{equation}
	H(\omega,\bm{\beta}) \approx H_{\rm geom}(\bm{\beta}) \equiv \sum_i |H_i(\bm{\beta})|.
	\label{eq:Hgeo}
\end{equation}
The validity of this approximation is determined by the angular scale of oscillations in the Eikonal magnification,
\begin{equation}
\Delta\theta_{\rm osc} \equiv \left| \frac{\omega}{2\pi} \bm{\nabla}_{\bm{\beta}}
	 \left\{ T(\bm{\theta}_j, \bm{\beta}) - T(\bm{\theta}_i, \bm{\beta}) \right\} \right|^{-1},
	 \label{eq:Deltathetaosc}
\end{equation}
and other angular scales in the problem, as well as the scale of oscillation in the frequency domain,
\begin{equation}
    \Delta \omega_{\rm osc} \equiv \left| \frac{1}{2\pi} \left\{ T(\bm{\theta}_j, \bm{\beta}) - T(\bm{\theta}_i, \bm{\beta}) \right\} \right|^{-1}.
\label{eq:wosc}
\end{equation}
In particular, the geometric limit holds if either of two conditions is met:
\begin{enumerate}
    \item when the effective observational resolution in $\beta$ is low
compared to $\Delta\theta_{\rm osc}$, and the frequency resolution is low compared to $\Delta \omega_{\rm osc}$. In the limit as $\omega \to \infty$, the second term in Eq.~\ref{eq:semiclassical} oscillates rapidly as a function of $\bm{\beta}$, and so if we compute the average magnification over a region of size $\Delta \beta$ around a given $\bm{\beta}$, the second term vanishes, and we are left with the geometric optics result as a function $\beta$. If, however, $\Delta \omega_{\rm osc}$ is comparable to the bandwidth of the observation and larger than the frequency resolution, then an interference pattern may still be observed in the frequency domain. The nominal Rayleigh ``resolution'' is typically not the limiting factor, but rather the time or frequency resolution can be limited when the source is too faint to detect individual scintles, requiring wide band averages that lead to an effective smearing in the observer plane.
    \item when the source is extended. The result in Eq.~\ref{eq:F} was obtained assuming a point-source of coherent radiation. When the size of the source becomes larger than the scale of the interference fringes ($\sim\Delta\theta_{\rm osc}$), then the phase coherence of the radiation is lost. When phase coherence is lost, the oscillations must be averaged over, and the geometric optics result is obtained. 
\end{enumerate}
For many cases of interest in gravitational lensing, the second of these conditions is met, so that geometric optics is sufficient for many astrophysical applications. In the following sections, we will describe situations where this is no longer true, and wave optics effects must be accounted for.

\subsection{Point lenses}
\label{sec:point_lens}

We now turn our attention to a specific application of the formalism we have introduced so far; namely, the application to single point lenses. Consider a point lens of mass $M$, with the optical axis chosen so that the mass is located at $\bm{\theta}=0$, and with the source at a fixed location $\bm{\beta}$ on the sky, i.e. with no relative motion with respect to the lens. The Fermat potential in this case is given by \citep{Schneider}
\begin{equation}
    T(\bm{\theta}, \bm{\beta}) = \frac{D}{2} |\bm{\theta} - \bm{\beta}|^2 - 4GM \log |\bm{\theta}|.
\end{equation}
This has two stationary points, located at
\begin{equation}
    \bm{x}_{\pm} = \frac{\bm{y}}{2y} \left(y \pm \sqrt{4 + y^2} \right),
    \label{eq:pnt_im_pos}
\end{equation}
with geometric optics magnifications of
\begin{align}
    H_{\pm} &= \frac{y^2 + 2}{2y \sqrt{y^2 + 4}} \pm \frac{1}{2} 
    \label{eq:pnt_12} \\
    H_\mathrm{geom} &= H_+ + H_- = \frac{y^2 + 2}{y \sqrt{y^2 + 4}}.
    \label{eq:pnt_geom}
\end{align}
Here we have defined the dimensionless quantities $\bm{y} = \bm{\beta}/\theta_E$ and $\bm{x} = \bm{\theta}/\theta_E$, where $\theta_E = \sqrt{4GM/D}$ is the Einstein radius. Note that the lens mapping has a singularity at $y=0$.

From Eq.~\ref{eq:semiclassical}, the magnification in the Eikonal, or semi-classical limit, is \citep{PhysRevD.34.1708, 1986ApJ...307...30D}
\begin{align}
\begin{split}
    H_\mathrm{Eik} &= \frac{y^2 + 2 + 2\sin(s\hat{T}_{12})}{y \sqrt{y^2 + 4}}, \\
    \hat{T}_{12} &= \frac{1}{2}y\sqrt{y^2+4} + \log \left( \frac{\sqrt{y^2+4}+y}{\sqrt{y^2+4}-y} \right),
\end{split}
    \label{eq:pnt_semi}
\end{align}
where
\begin{equation}
s = 4GM\omega = \left( \frac{\theta_E}{\theta_F} \right)^2.
\end{equation}
Note, this quantity is the square of the ratio of the two fundamental angular scales in the problem: the Einstein radius and the Fresnel scale. This ratio will play a central role in indicating the importance of wave effects. The Eikonal approximation is valid when $\beta \gg \theta_F$, i.e. when $y \gg y_F \equiv \theta_F/\theta_E = \sqrt{1/s}$. This is roughly because as $\beta$ increases beyond the Fresnel scale, the regions around the two geometric images that contribute to the diffraction integral become well separated. 

We can also define $s$ using physical scales instead of angular scales: since $\theta_E = \sqrt{2r_{\rm s}/D}$ where $r_{\rm s}$ is the Schwarzschild radius of the lens and $\theta_F=\sqrt{\lambda/2\pi D}$, we can write
\begin{equation}
	s = 4\pi\frac{r_{\rm s}}{\lambda}\ .
\end{equation}
Thus, wave effects become important when the Schwarzschild radius of the lens is comparable to the wavelength of the incident light, a fact that has previously been found to be relevant for lensing of gravitational waves \citep[e.g.][]{Takahashi:2003ix} and gamma-ray bursts \citep[e.g.][]{Gould:1992}.

The scale of the oscillations in $\beta$ in the Eikonal limit is given by Eq.~\ref{eq:Deltathetaosc}, and is computed for the point lens to be
\begin{equation}
    \Delta \theta_{\rm osc} = 2\pi \frac{\theta^2_F}{\theta_E} \sqrt{4 + \left(\frac{\beta}{\theta_E}\right)^2} \sim 4\pi \frac{\theta_F}{\sqrt{s}}.
    \label{eq:pntthetaosc}
\end{equation}
The full evaluation of the integral in Eq.~\ref{eq:F} for the point lens gives \citep{nakamura}
\begin{equation}
    H_\mathrm{wave} = |F|^2 = \frac{\pi s}{1 - e^{-\pi s}} |{}_1 F_1 (\frac{1}{2} is, 1; \frac{1}{2} isy^2) |^2,
    \label{eq:Hwave}
\end{equation}
where ${}_1 F_1$ is a confluent hypergeometric function. Thus, the magnification of a source at a fixed position on the sky depends on two parameters: the source's angular separation from the lens in units of Einstein radii, and the ratio of the Einstein radius to the Fresnel scale. In the geometric optics limit [Eq.~\eqref{eq:Hgeo}], the magnification depends on only the first of these parameters. The wave optics result is therefore chromatic, unlike the geometric optics result. We explore these parameter dependencies further in the next section.

For many astrophysical applications of lensing, the sources in question are larger than the scale of oscillations in the diffraction pattern given by Eq.~\ref{eq:pntthetaosc}, and so for most cases of interest the geometric limit of optics is sufficient. For example, the microlensing of background stars in the Milky Way has yielded a fruitful method of detecting exoplanets in orbit around stars \citep[see e.g.][]{gaudi_microlensing_2012,Schneider}. In this case, the background stars have typical angular size $\sim 1\,\mu$as, while $\Delta \theta_\mathrm{osc}\lesssim 10^{-3}\,\mu$as, placing this situation firmly in the geometric limit. In contrast, both pulsars and FRBs are indeed effectively point sources of coherent radiation \citep{2016arXiv160207738K, Petroff2019}. Thus, wave optical effects will generically be important in the gravitational lensing of these sources. 

\section{Parameter Inference}
\label{sec:param_inference}

The chromatic nature of diffractive effects in a lensing event can in principle be used to extract more information about the lens and source than would be possible in the case of geometric optics. In this section, we demonstrate this explicitly for the situation of a point lens and point source.

\subsection{Point lens and point source in relative motion}
\label{sec:rel_motion}

As in Section~\ref{sec:point_lens}, we continue to consider the specific application of wave optics in the case of a point source lensed by a single point mass. In general, astrophysical sources and lenses are not stationary on the sky, but move relative to each other. Suppose that the lens and source are moving with a relative angular velocity $\bm{\mu}_\mathrm{rel}$, measured in terms of the change in relative angular position of the source and lens on the sky per unit time. As a result, the source will have a magnification that depends on time. Assuming that over the period of interest, the source and lens move with constant relative velocity, the position of the angular distance of the source from the lens (normalized by $\theta_E$) will be given by
\begin{equation}
    y(\tau) = (\tau^2 + y_0^2)^{1/2},
\end{equation}
where $\tau \equiv (t-t_0)/t_E$, and $t_E \equiv \theta_E/ \mu_\mathrm{rel}$ is the Einstein radius crossing time. Thus, the magnification, given by Eq.~\ref{eq:Hwave}, will have a time dependence since $y$ varies with time. We have set $y(0) = y_0$ to be the value of $y$ at closest approach, and $t_0$ to be the time at which this occurs. In general, for microlensing events, the source being lensed is not observed constantly throughout the entire duration of the event, and so the time of peak brightness, $t_0$, needs to be fit as well.

The full set of parameters that the wave optics magnification depends on is then $(t, s, t_E, y_0, t_0)$. Now, since the parameter $s$ is frequency dependent, and what is actually observed in an experiment is a dynamic spectrum (i.e. a set of light curves for a range of frequencies), we can further decompose $s$ into its frequency dependent and independent parts. Since $s \propto M\omega$ we can parameterize the wave optics magnification as $H_\mathrm{wave}(t, \omega ; M, t_E, y_0, t_0)$. Thus, the dynamic spectrum is fully specified by the parameters $(M, t_E, y_0, t_0)$.\footnote{ Note that it is possible to reparametrize in terms of the Fresnel crossing time, $t_F = \theta_F/\mu_{\rm rel.}$, and the impact parameter $z_0 = \beta_0/\theta_F$; however, these are frequency dependent quantities, and so it is more convenient to parametrize the dynamic spectrum by the frequency independent quantities $t_E$ and $y_0$.} One could also define a set of parameters involving the underlying physical parameters describing the lensing geometry: then $(M, t_E, y_0, t_0)$ becomes $(M, D, \mu_{\rm rel}, \beta_0, t_0)$. However, there is a degeneracy between $D$ and $\mu_\mathrm{rel}$, since the dependence on these parameters only enters through the fact that $t_E \sim (M/\mu_\mathrm{rel}^2 D)^{1/2}$. Contrast this with the geometric optics case, where the light curves have no frequency dependence, and are completely specified by $(t_E, y_0, t_0)$. In this case, there is a threeway degeneracy between $M$, $D$, and $\mu_\mathrm{rel}$, since the mass $M$ also only enters through $t_E$. Thus, while in the geometric case, the mass is degenerate with two other physical parameters, wave optics effects break this degeneracy. 

\begin{figure*}
    \centering
    \includegraphics[width=2\columnwidth]{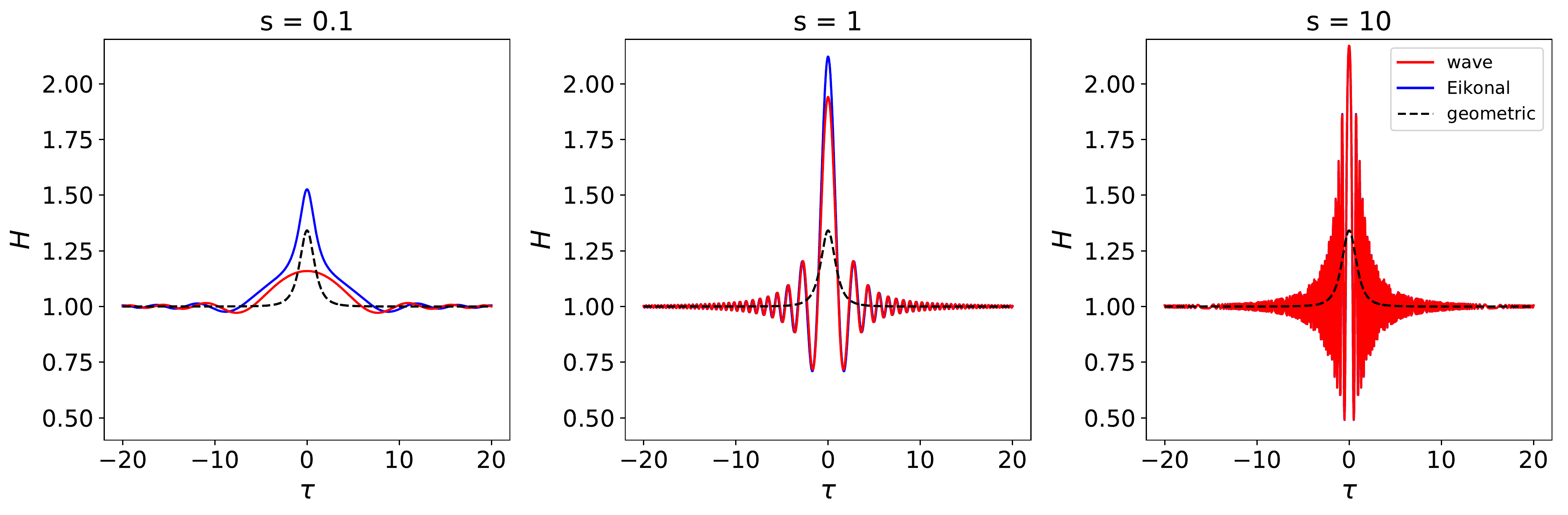}
    \caption{
    Magnification curves of a point source lensed by a point mass in relative motion with the source, as a function of the dimensionless time $\tau = t/t_E$, where $t_E$ is the time it takes for the source to move one Einstein radius relative to the lens and $\tau=0$ corresponds to the point of closest approach between the source and lens. The red line is the magnification computed in the full wave-optics regime, and the blue and black curves are the magnification evaluated in the Eikonal and geometric limits, respectively. The curves are plotted for three different values of $s$. Note that as $s$ increases (i.e. as $\theta_E$ becomes much larger than $\theta_F$), the Eikonal result better approximates the wave optics result, and that by averaging over the oscillations one obtains the geometric result which does not depend on $s$.}
    \label{fig:schrwz_lcs}
\end{figure*}

Fig.~\ref{fig:schrwz_lcs} shows a comparison of the light curves as a function of $\tau$, computed in the full wave-optics regime, the Eikonal limit, and the geometric limit, for fixed frequencies. The light curves are computed for $y_0 = 1$, and for $s = 0.1, 1, 10$. We see that the Eikonal limit is only a valid approximation for large $s$. This happens because the Eikonal limit is only valid when the two geometric images are well separated compared to the Fresnel scale. The scale of the separation of the two images is the Einstein radius; as $s$ increases, the Fresnel scale becomes small compared to the Einstein radius, and so the images become well-separated. Note that the Eikonal approximation will always be worse near $\tau = 0$, since $\tau=0$ corresponds to the minimum separation between source and lens, and therefore, by Eq.~\ref{eq:pnt_im_pos}, the minimum separation between images. 
\begin{figure*}
    \centering
    \includegraphics[width=2\columnwidth]{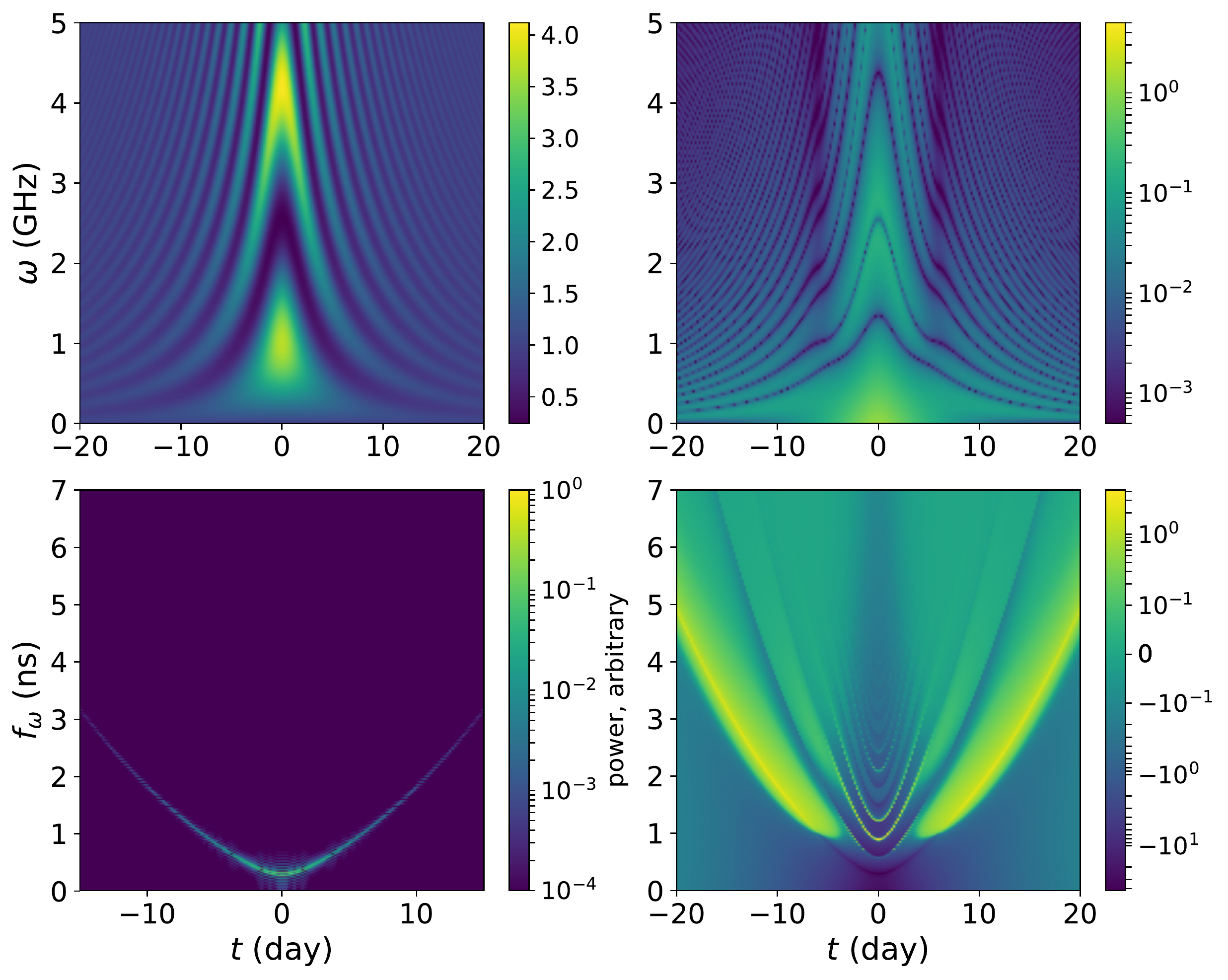}
    \caption{(Top left) Dynamic spectrum of a point source lensed by a point mass given by the parameters $M = 5\,M_\oplus$, $D = 1\,$kpc, $\mu_\mathrm{rel} = 1\,$mas/year, $y_0 = 0.5$, and $t_0 = 0$. The colour gives the magnification of the point source. (Top right) The fractional difference between the full wave optics result given by Eq.~\ref{eq:Hwave} and the Eikonal limit given by Eq.~\ref{eq:pnt_semi}. 
    (Bottom left) The top panel Fourier transformed along the frequency axis. In the Eikonal limit, the 1D Fourier transform shown in the bottom plot only has non-zero power when $f_\omega$ is equal to the time delay between the two images at time $t$, which can be used to read off the instantaneous time delay. (Bottom right) The real part of the Fourier transform along the frequency axis of the fractional difference between the full wave optics result and the Eikonal limit.
    }
    \label{fig:spect}
\end{figure*}

The top left panel of Fig.~\ref{fig:spect} shows the dynamic spectrum, i.e. the magnification as a function of frequency and time, of the full wave-optics result computed from Eq.~\ref{eq:Hwave}, with physical parameters $M=5\,M_\oplus$, $D=1\,$kpc, $\mu_\mathrm{rel} = 10\,$mas/year, $y_0 = 0.5$, and $t_0 = 0$, which correspond to an Einstein crossing time of $t_E = 0.4\,$days. We have chosen values of the parameters that are typical in planetary microlensing~\citep{gaudi_microlensing_2012}, which we discuss further in Section~\ref{sec:planets}. For the central frequency, $\omega = 2\,$GHz, these parameters give $s = 3.6$. The top right panel of Fig.~\ref{fig:spect} shows the fractional deviation of the full wave result from the Eikonal limit, and the bottom panel shows the dynamic spectrum with the Fourier transform performed along the frequency axis. As the Eikonal approximation becomes valid (i.e. as $|t|$ increases), the magnification as a function of frequency, $\omega$, for fixed $t$ becomes a simple sinusoid with frequency $4GM \hat{T}_{12}$, which can easily be seen from Eq.~\ref{eq:pnt_semi}. Likewise, the Fourier transform along the frequency axis, shown in the bottom left panel of Fig.~\ref{fig:spect}, becomes a delta function centred at $4GM \hat{T}_{12}$, where this quantity is just the dimensionful value of the time delay between the two geometric images. Thus, from this 1D Fourier transform we can read off an instantaneous measurement of the time delay between the two images. The bottom right panel of Fig.~\ref{fig:spect} shows the Fourier transform along the frequency axis of the fractional difference between the full wave optics result and the Eikonal limit. Defining the magnification to be equal for positive and negative frequencies, the imaginary part of the Fourier transform vanishes and the only real part is shown. The 1D Fourier transform has features at the parabolas defined by integer multiples of the quantity $4GM \hat{T}_{12}$.

\begin{figure}
    \centering
    \includegraphics[width=\columnwidth]{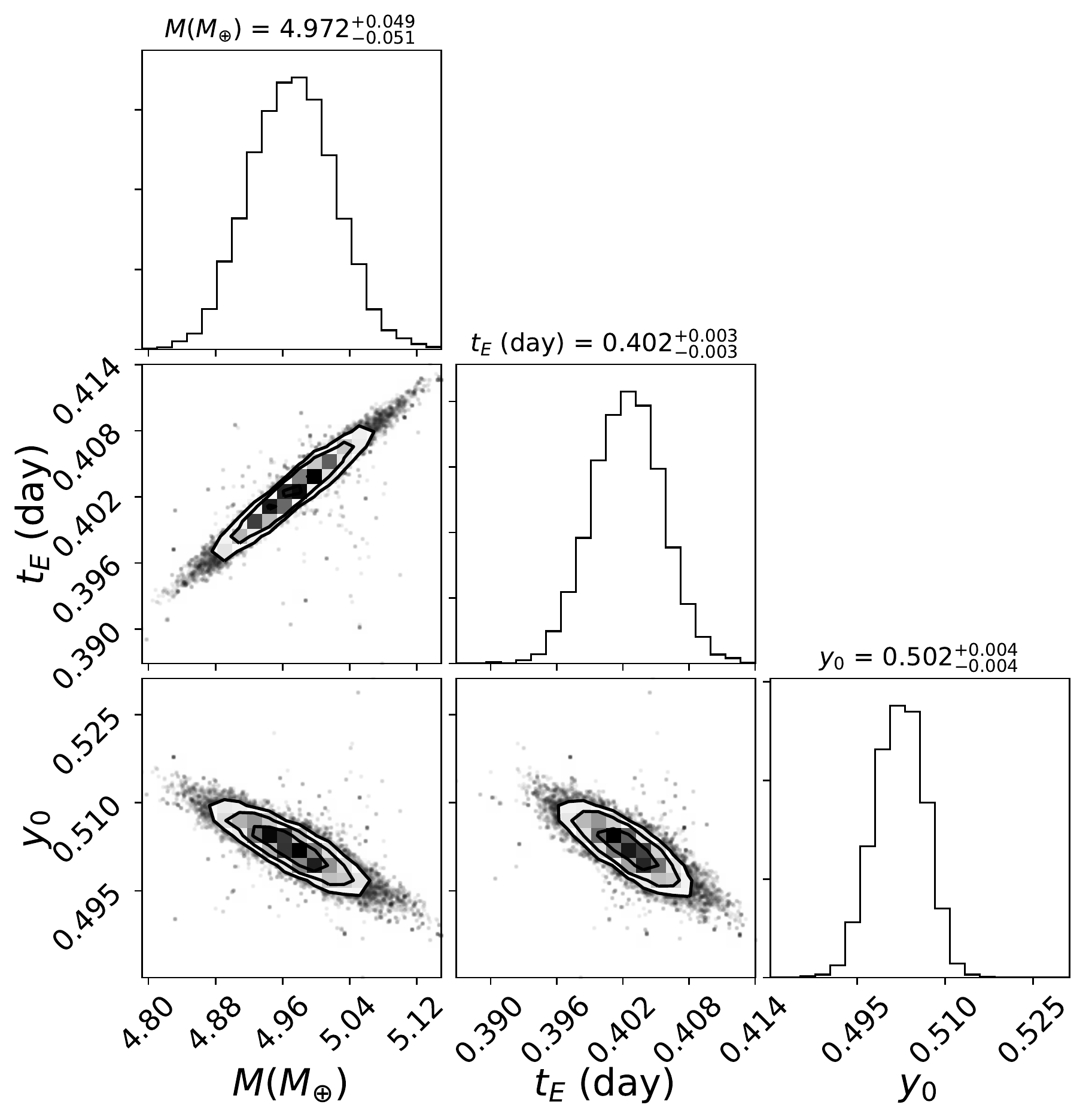}
    \caption{Result of a simple parameter estimation using MCMC on simulated data, shown in Fig.~\ref{fig:spect}, of a point source lensed by a point mass in the wave-optics regime. The data are a set of light curves computed from $H_\mathrm{wave}(t, \omega; M, t_E, y_0)$, as a function of time, for a set of frequencies in the range of $1-2\,$GHz. Physical parameters of $M = 5\, M_\oplus$, $D = 1\,$kpc, $\mu_\mathrm{rel} = 10\,$mas/year, and $y_0 = 0.5$ were chosen, corresponding to $t_E = 0.4\,$days. We generated the light curves over a long enough period of time so that 3 diffraction peaks were present on either side of the central peak, and Gaussian noise was added to the magnification of constant $\sigma = 0.1$. From the simulated data and choosing log-uniform priors on $M$ and $t_E$, and a uniform prior on $y_0$, we were able to infer the parameters with high signal-to-noise. Inference of the mass would be impossible for a light curve following the geometric optics result, as the geometric result depends only on $y_0$ and $t_E = \theta_E / \mu_\mathrm{rel}$, resulting in a total degeneracy between $M$, $D$, and $\mu_\mathrm{rel}$.}
    \label{fig:MCMC}
\end{figure}

Fig.~\ref{fig:MCMC} shows the results of using MCMC to infer the underlying parameters of a set of simulated data. The simulated data was the dynamic spectrum shown in Fig.~\ref{fig:spect} with Gaussian noise added per time/frequency sample, with $\sigma = 0.1$. For the MCMC inference, we chose log-uniform priors on $M$ and $t_E$, and a uniform prior on $y_0$. Here, for simplicity, we assume the light curve is well sampled along the entire duration, and so we fix $t_0 = 0$. Fig.~\ref{fig:MCMC} shows that we are able to place tight constraints on all of the parameters, including the mass\footnote{The MCMC analysis was carried out using the \texttt{python} package \texttt{emcee} \citep{emcee}, with 50 walkers and 500 steps.}. Thus, wave optical effects do indeed break the degeneracy between the mass and the Einstein crossing time that is present in the geometric regime. While in the geometric case only the Einstein crossing time and the parameter $y_0$ were accessible from a measurement of the light curve, through the observation of diffractive effects, the mass of the lens also becomes eminently measurable. 

\subsection{Static source and lens}
\label{sec:static}

If the lens and source are stationary relative to each other, is it still possible to measure the mass of the lens in the wave-optics regime? Such a situation might occur in a microlensing event of an FRB, which are bursts of millisecond duration that irregularly repeat, when they repeat at all. For microlensing events that have event durations much longer than a millisecond, such as microlensing by planets in the galaxy (see Section~\ref{sec:applications}), a single millisecond burst is not long enough to sample the dynamic spectrum for the entire event. Instead, such an observation would provide only a single time slice of the dynamic spectrum. 

Obviously, a spectrum at a single instant in time provides much less constraining power than the full dynamic spectrum; nonetheless, it is still possible to infer the mass. This is easiest to see in the Eikonal limit using Eq.~\ref{eq:pnt_semi}, however, the same conclusion holds in the full wave-optics regime. In Section~\ref{sec:rel_motion} we saw that a measurement of the magnification as a function of frequency at a single time slice gives a direct measurement of the time delay $4GM\hat{T}_{12}$, where $\hat{T}_{12}$ only depends on $y$. Thus, if we could measure the separation $y$ between the source and lens, then we can instantaneously measure the mass of the lens. The amplitude of the oscillations as a function of frequency give us this measurement of $y$. In the Eikonal limit, the height of the peaks in the magnification is given by
\begin{equation}
    \max_s \left[H_\mathrm{Eik}(s,y)\right] = \frac{\sqrt{y^2+4}}{y},
\end{equation}
which can be inverted to determine $y$. Thus, an observation of the dynamic spectrum at a single time slice is sufficient to estimate the mass of the lens.

Note that in this section we have assumed the dynamic spectrum of the source is known so that the magnification of the source due to lensing can be disentangled from intrinsic changes due to the variation in the source itself. In the case of pulsars as the background source, there is significant variation between individual pulses; however, the average profile over many pulses is highly stable. Thus, variations in this average profile will effectively be equal to the magnification due to lensing. For more general sources, an appropriate threshold on the magnification for a given event needs to be chosen so that the change in brightness is due mostly to lensing and not intrinsic variation. An advantage of wave-optical effects is that they allow for an instantaneous measurement of the mass, and so a source may be allowed to significantly vary in brightness over time, so long as its brightness as a function of frequency is constant. In the case of FRBs, for which the brightness often varies significantly between bursts (for repeating FRBs), the brightness is typically more stable as a function of frequency.

It is also important to note that Eq.~\ref{eq:Hwave} provides a template for the modulation of the dynamic spectrum of a source by a point lens. Thus, detection of a lensing event will not be limited by the signal-to-noise of any given diffraction peak. By matching observations against a template, one can in principle achieve much higher sensitivities than would otherwise be possible.

\section{Cross sections}
\label{sec:crssxn}

\begin{figure}
    \centering
    \includegraphics[width=\columnwidth]{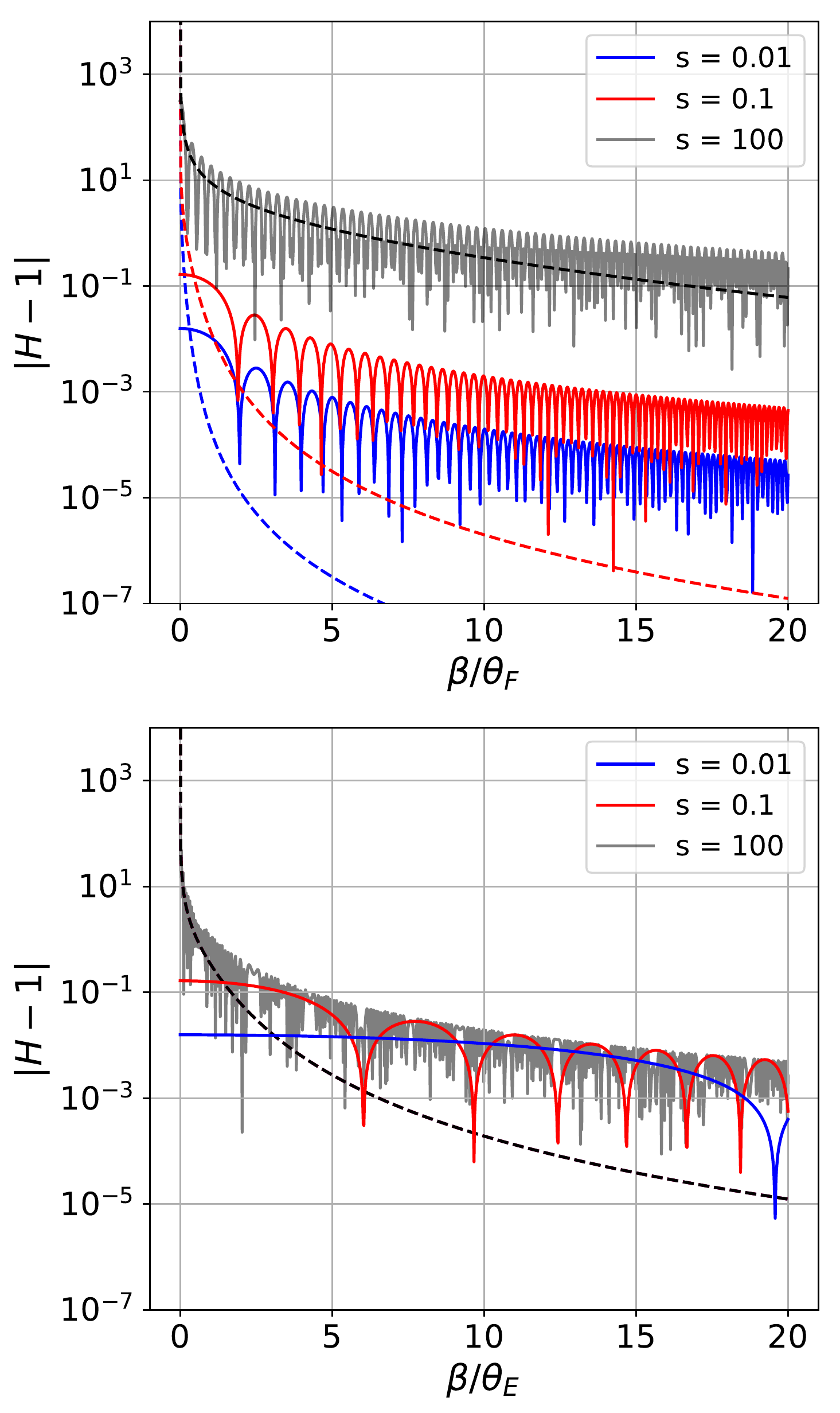}
    \caption{
    The magnification of a point source lensed by a point mass as a function of the angular position of the source relative to the lens, normalized by the Fresnel scale, $\theta_F$, in the top panel, and normalized by the Einstein radius, $\theta_E$, in the bottom panel. The $y$-axis shows the absolute value of the difference of the magnification from one (the value of $H$ in the absence of the lens). The dashed curves show the result for the geometric limit, whereas the solid curves show the result in the full wave-optics regime. The curves are plotted for three different values of $s$. Note that in the right panel, all of the dashed lines overlap, since the geometric result as a function of $\beta/\theta_E$ is independent of $s$. For high $s$, the average of the wave-optics curve over its oscillations approaches the geometric result. However, the amplitude of the wave optics results decays much slower than the geometric result, and so as the diffraction peaks become resolvable, the effective microlensing cross-section is increased relative to the geometric cross-section.}
    \label{fig:frnl}
\end{figure}

So far, we have presented results for the wave-optical effects of a point lens that can be found elsewhere in the literature. We have attempted to clearly lay out the conditions for the different regimes, and we have emphasized  that a measurement of a gravitational lensing event in the wave-optics regime can break degeneracies that are present in the geometric regime, due to the frequency dependence of wave-optical effects. But in order to assess whether or not such events are likely to occur, we will also need to compute the cross section for a given event, i.e. the measure of the set of points in the source plane where the source can be located for some effect to be observed. Often one defines the cross section for a microlensing event in the geometric optics regime to be the area of source positions that give rise to a magnification above some threshold. In the wave-optics regime, one might similarly define the cross section to be the region in the source plane where the \textit{amplitude of the oscillations} in the magnification is larger than some threshold.  Fig.~\ref{fig:frnl} shows a comparison of the geometric and wave-optical magnifications of a point source by a point lens as a function of source position, $\beta$, normalized in the top panel by the Fresnel scale, $\theta_F$, and normalized in the bottom panel by the Einstein radius, $\theta_E$. The figure shows that, independent of the value of $s$, the amplitude of oscillations in the wave-optics magnification decays more slowly with $\beta$ than the value of the geometric magnification. Thus, when the diffraction peaks are resolvable, the effective cross section for lensing is larger than it would be in the geometric limit.

To obtain expressions for the cross section in different cases, we consider it to be defined by an amplification threshold $A_H$. In the geometric case, this cross section is determined by the value of $y_*$ for which $|H_\mathrm{geom}(y_*)-1|=A_H$. Since $H_\mathrm{geom}(y)$ is a decreasing function, $|H_\mathrm{geom}(y)-1|>A_H$ whenever $y < y_*$, and so the angular cross section is $\sigma = \pi y_*^2 \theta_E^2$. We will typically be sensitive to changes in brightness well below order unity, so $A_H \ll 1$. As a result, the typical values of $y_*$ will be greater than 1, and so to obtain an easily invertible expression for $y_*$ we can Taylor expand for large $y$. To first non-trivial order, the geometric magnification (Eq.~\ref{eq:pnt_geom}) is
\begin{equation}
    H_\mathrm{geom}(y) \approx 1 + \frac{2}{y^4} + \mathcal{O}\left( \frac{1}{y^6} \right).
\end{equation}
Given a threshold $A_H$, we then have
\begin{equation}
y_* \approx \left( \frac{2}{A_H} \right)^{1/4},
\end{equation}
implying that
\begin{equation}
\sigma_\mathrm{geom} = \pi y_*^2 \theta^2_E \approx \pi \sqrt{\frac{2}{A_H}} \theta_E^2.
\end{equation}

In the wave optics regime, the cross section will depend on how the geometric cross section compares to the area defined by the Fresnel scale, i.e.\ how $y_*$ compares to $y_F \equiv \theta_F/\theta_E$. To investigate the case when $y_* \gtrsim y_F$, first recall that the Eikonal approximation is valid when $y\gg y_F$. The Taylor expansion of the Eikonal magnification in Eq.~\eqref{eq:pnt_semi} for large $y$ is given by
\begin{equation}
    H_{\rm Eik}(y, s) \approx 1 + \frac{2\sin(s\hat{T}_{12})}{y^2} + \mathcal{O}\left( \frac{1}{y^4}\right).
\end{equation}
Thus, the amplitude of the magnification above unity in the Eikonal approximation decays as $2/y^2$, much more slowly than the $2/y^4$ decay in the geometric limit. This implies that if $y_* \gtrsim y_F$, then the lensing cross section will be larger than the geometric result. Specifically, the condition $|H_\mathrm{Eik}(\tilde{y})-1|=A_H$ yields $\tilde{y} \approx (2/A_H)^{1/2}$, and a cross section of 
\begin{equation}
    \sigma_\mathrm{wave} = \pi \tilde{y}^2 \theta_E^2 = \pi y_*^4 \theta^2_E \approx \pi \frac{2}{A_H} \theta_E^2.
    \label{eq:sigwave-large}
\end{equation}

On the other hand, when $y_* \ll y_F$, we no longer have that $\tilde{y} \gg y_F$, and so we cannot use the Eikonal approximation. In this regime, the wave optics magnification drops below the threshold at a $y$ value too low for the Eikonal limit to be valid. This happens after a range of $y$ covering only a few oscillations of the magnification. Recall that the scale of these oscillations is given by $4\pi \theta_F/\sqrt{s}$ (Eq.~\ref{eq:pntthetaosc}), and so we can roughly estimate the cross section in this case as
\begin{equation}
\sigma_{\rm wave} \sim \pi \Delta \theta_{\rm osc}^2 = \frac{16\pi^3}{s} \theta^2_F
	= \frac{16\pi^3}{s^2} \theta_E^2\ .
\end{equation}
It is also possible that the amplitude never rises above the threshold. This happens when $|H_\mathrm{wave}(0, s) - 1| < A_H$. In this case, the cross section is simply zero. 

\begin{table*}
\begin{center}
    \begin{tabular}{ | c |  l |}
    \hline
    $\sigma$ & Condition  \\ \hline 
    $\sqrt{\frac{2}{A_H}} \times \pi  \theta_E^2$ & Geometric optics \\
    $\frac{2}{A_H} \times \pi  \theta_E^2$ & Wave optics, when $\sigma_{\rm geom} \gtrsim \pi\theta_F^2$  \\
    $\frac{16\pi^2}{s^2} \times \pi \theta_E^2$ & Wave optics, when $\sigma_{\rm geom} \lesssim \pi\theta_F^2$ \\
    $0$ & Wave optics; maximum magnification always below threshold
    \\ \hline
    \end{tabular}
    \caption{
    Cross sections for microlensing in different regimes, defined by the region in the source plane for which the fractional magnification of the source exceeds a threshold $A_H$. In all cases with nonzero cross section, wave-optical effects act to strongly enhance the cross section over the geometric optics result.
    }
    \label{tab:crsxn}
\end{center}
\end{table*}

Table~\ref{tab:crsxn} summarizes the cross sections in the different regimes discussed above. In all cases with nonzero cross section (i.e.\ for which the magnification exceeds $A_H$ at some source position), we find that the wave optics result is strongly boosted with respect to the geometric optics cross section. (We will continue to assume that $A_H<1$.) When $\sigma_{\rm geom} \gtrsim \pi\theta_F^2$, $\sigma_{\rm wave}$ exceeds $\sigma_{\rm geom}$ by a factor $(2/A_H)^{1/2}$; for a percent-level threshold, this implies an order-of-magnitude enhancement of the cross section. When $\sigma_{\rm geom} \lesssim \pi\theta_F^2$, there is an enhancement when $s\lesssim (8A_H)^{1/4} \pi \approx 5.3A_H^{1/4}$. The only time the geometric cross section is larger than the wave-optics result is when $s \gtrsim 5.3A_H^{1/4}$ and we have chosen a magnification threshold close to the maximum magnification, $H_\mathrm{wave}(0,s)$, for that value of $s$.

It may seem, at first glance, that the geometric result is always larger than the wave-optics result for the case where $\sigma_\mathrm{wave} = 0$; however, this is merely an artefact of the formal infinity in the geometric magnification. In reality, the magnification never exceeds $H_\mathrm{wave}(0, s)$. \citet{Montero-Camacho:2019jte} find that this fictitious increase in the geometric cross section compared to the wave optics cross sections leads to overestimates of event rates in the microlensing of stars by PBHs. Thus, in a wide variety of cases, wave-optical effects will lead to an effective boost in the microlensing cross section. This is consistent with the finding of \citet{Takahashi:2003ix} that the lens mass could be constrained with higher signal-to-noise in the lensing of gravitational waves for larger source distances.

\section{Applications of the point-lens model}
\label{sec:applications}

We have so far argued that microlensing events in the wave-optics regime have two distinct advantages over events in the geometric regime: they contain more information about the lensing system, and they generically have an increased cross section. In the particular example of a point-mass lens, we have shown that an observation of the interference pattern can be used to break the degeneracy between the mass of the lens and the Einstein crossing time, $t_E$. We now investigate what kinds of point lenses and sources we might be able to observe in the wave optics regime.

\subsection{Relevant scales}
\label{sec:scales}

As we have seen, in the wave-optics regime of a point lens, there are two relevant angular scales: the Fresnel scale, $\theta_F$, and the Einstein radius, $\theta_E$. Here, we give the values of the these scales in terms of the physical parameters, normalized to typical values that will be useful in the following discussion. The Fresnel scale is given by
\begin{equation}
    \theta_F = \sqrt{\frac{1}{\omega D}} = 8.1\,\mu\mathrm{as}\,\left(\frac{\nu}{\rm GHz}\right)^{-1/2} \left(\frac{D}{\rm kpc}\right)^{-1/2},
    \label{eq:num_thf}
\end{equation}
where $\omega = 2\pi \nu$, and the angular Einstein radius is
\begin{equation}
    \theta_E = \sqrt{\frac{4GM}{D}} = 4.9\,\mu{\rm as} \left(\frac{M}{M_\oplus} \right)^{1/2} \left(\frac{D}{\rm kpc} \right)^{-1/2}.
    \label{eq:num_the}
\end{equation}
The square of the ratio between the two scales, $s = (\theta_E/\theta_F)^2$, which, heuristically, sets the relative importance of geometric effects to wave effects is given by
\begin{equation}
    s = 4GM\omega = 0.36\,\left(\frac{M}{ M_\oplus}\right)\left(\frac{\nu}{\rm GHz}\right).
    \label{eq:num_s}
\end{equation}

When the lens and source are moving relative to each other with angular velocity $\mu_{\rm rel.}$, the time scales $t_F$ and $t_E$ become relevant. The scale that determines the spacing of the diffraction peaks as a function of $\beta$ is given by Eq.~\eqref{eq:pntthetaosc}:
\begin{equation}
\Delta \theta_{\rm osc} = 4\pi \frac{\theta_F}{\sqrt{s}}
	= 0.17\,{\rm mas} \left( \frac{M}{ M_\oplus} \right)^{1/2} \left(\frac{D}{\rm kpc} \right)^{-1/2}
	\left(\frac{\nu}{\rm GHz}\right)^{-1}.
\end{equation}
For a microlensing event where source and lens are in relative motion, in order to observe the diffraction peaks as a function of time in the dynamic spectrum, one must observe the lensing event for at least the time it takes the source and lens to move $\Delta \theta_{\rm osc}$ relative to each other. The Fresnel crossing time is given by
\begin{equation}
    t_F = \frac{\theta_F}{\mu_\mathrm{rel}} = 23.2\,\mathrm{day} \left(\frac{\mu_{\rm rel.}}{\mathrm{mas}/\mathrm{year}} \right)^{-1} \left(\frac{\nu}{\rm GHz}\right)^{-1/2} \left(\frac{D}{\rm kpc}\right)^{-1/2}.
    \label{eq:num_tf}
\end{equation}
and the time it takes to cross $\Delta \theta_{\rm osc}$ is $4\pi t_F/\sqrt{s}$. For a broad range of $s$ around 1, the time scale for observing oscillations is roughly the Fresnel crossing time.

The total duration of the event depends on $s$. When $s$ is greater than one, the Eikonal limit becomes a good approximation, and so the time scale for the event is given by the Einstein crossing time,
\begin{equation}
    t_E = \frac{\theta_E}{\mu_\mathrm{rel}} =1.7\,\mathrm{day} \left(\frac{\mu_{\rm rel.}}{\mathrm{mas}/\mathrm{year}} \right)^{-1}\left(\frac{M}{M_\oplus}\right)^{1/2}\left(\frac{D}{\rm kpc}\right)^{-1/2}.
    \label{eq:num_te}
\end{equation}
When $s$ is less than one, the duration of the event is roughly the width of the first diffraction peak (see Fig.~\ref{fig:schrwz_lcs}), and so $t_F/\sqrt{s}$ sets the duration. 

\subsection{Microlensing by free-floating planets}
\label{sec:planets}

In this section we will consider the case when $s \lesssim 1$, so that both the Eikonal and geometric approximations are poor approximations of the relative motion light curve as a function of $\tau$. That is, we want to know what kind of lenses will lens a coherent source in the full wave-optics regime. In order for the full wave-optics regime to be observable in the point-lens case, additional criteria need to be satisfied: the size of the source needs to be less than the Fresnel scale (the source needs to be a point source), and we will also require that the maximum magnification of the source be greater than some threshold. The maximum magnification occurs when source and lens are aligned, i.e. when $y=0$. In geometric optics, this is a caustic, and so the magnification is infinite. However, in the wave-optics regime, the magnification is bounded by
\begin{equation}
    \max_y | H_{\rm wave}(s, y)-1 |  = |H_{\rm wave} (s, 0) - 1| = \frac{\pi s}{1 - e^{-\pi s}} - 1.
\end{equation}
The condition that this be larger than some threshold is just the condition that the lens be strong enough to measurably magnify the background source. This condition will place a lower bound on the relevant values of $s$. Taking a 10\% magnification as a rough benchmark for what is observable, i.e. requiring that $\max_y | H_{\rm wave}(s, y)-1 | \gtrsim 10^{-1}$, gives the condition that $s \gtrsim 0.06$. For coherent radio sources, such as pulsars and FRBs, and considering frequencies $\nu \sim 1\,$GHz, the requirement that $0.06 \lesssim s \lesssim 1$ becomes a condition on the relevant mass range: $M \in [0.1, 100]\,M_\oplus$. Therefore, point lenses of roughly planetary mass will significantly lens coherent radio sources in the full wave-optics regime. Note that this statement is independent of the distance to the lens or source, since this comes from constraints on $s$, which is independent of the lensing geometry. 

Objects that are isolated and fit this mass range include free-floating planets (FFPs). The planets need to be free-floating because lensing by planets bound to a stellar system will have effects from the host stars, requiring a lens model beyond the simple point lens we have been considering. Planet formation models predict that some fraction of planet forming stars will eject a number of their planets, resulting in a population of FFPs in the galaxy \citep{Ma2016}. Microlensing surveys have found distributions of the Einstein crossing time, $t_E$, associated with observed microlensing events to be consistent with a population of FFPs; although, the number and mass distribution of such objects is unclear \citep{Sumi2011, Mroz2017, 2018AJ....155..121M}. 

We also have the condition, however, that the sources be smaller than the Fresnel scale. For this we need to specify a distance scale. Let us consider lenses in our galaxy. The majority of lenses will be located in the galactic centre, so that the typical distance scale will be $D \sim 8\,kpc$. For radio sources at $\nu \sim 1\,$GHz, the Fresnel scale is $\theta_F \sim \mu$as. Thus, for a coherent source of radio emission to be effectively a point source, it must have angular dimensions much less than a microarcsecond. Both pulsars and FRBs sastify this condition, and so will potentially be lensed in the wave-optics regime by FFPs in the Milky Way. Even for extragalctic lenses, for which $D \sim 1$\,Gpc, FRBs, which are extragalactic sources, are much smaller than the corresponding Fresnel scale, $\theta_F \sim 10^{-3}\,\mu$as. In the rest of this section we will restrict our attention to lenses in the Milky Way, but will return to the question of extragalactic lenses in Section~\ref{sec:exgal_lenses}.

\subsubsection{Optical depth of FFPs in the Milky Way}
\label{sec:optical_depth}

We now want to estimate the optical depth for the lensing events described above; in other words, we want to estimate the probability that a given source is gravitationally lensed by an FFP in the galaxy in the relevant mass range, $[0.1, 100]\,M_\oplus$. Since the earth is located in a spiral arm of the Milky Way, the majority of these planets will be located towards the galactic centre. Thus, for simplicity, we will only consider lines of sight towards the galactic centre. If we require that that the source is magnified by at least ten percent ($A_H=10^{-1}$), such events will generally satisfy $\sigma_{\rm geom} \gtrsim \pi\theta_F^2$, and the wave-optics cross section will be given by Eq.~\eqref{eq:sigwave-large}. Note that this is larger than $\sigma_{\rm geom}$ by about a factor of 5.

The optical depth along a given line of sight up to a distance $D_s$ is given by
\begin{equation}
    \tau(D_s; l, b) = \int_0^{D_s} dD_d \int dM n(D_d, M; l, b) \sigma_{\rm wave}(D_d, D_s, M) D^2_d ,
    \label{eq:odepth}
\end{equation}
where $n(D_d, M; l, b)$ is the number density of free-floating planets as a function of mass and distance along the line of sight, and $(l, b)$ denotes the direction of the line of sight in galactic coordinates. Thus, in order to compute the optical depth we need to know the number and mass distribution of FFPs in the galaxy. This distribution, however, is not well-constrained. Core accretion models of planet formation predict a number of free-floating planets less than, but comparable to the number of stars \citep{Ma2016}. The mean mass of these planets is predicted to be of order earth mass, $\sim 1\,M_\oplus$. However, contrary to the theoretical prediction, \citet{Sumi2011} claim that a population of Jupiter-mass objects that are twice as numerous as stellar objects are required to explain their microlensing observations. Further observations later showed that the number of Jupiter-mass FFPs are more in line with the theoretical predictions \citep{Mroz2017}. Here we will assume that the number of FFPs is comparable to the number of stars, so that the number density of FFPs along any given line of sight, integrated over the mass, is equal to the number density of stars. We will also make the simplification, since we are only interested in obtaining a rough order-of-magnitude estimate of the optical depth and since the mass distribution of FFPs is observationally unknown, that all FFPs have a mass of $1\,M_\oplus$. Thus, we take the density in Eq.~\ref{eq:odepth} to be $n(D_d, M) = n_\mathrm{star}(D_d) \delta(M - 1\,M_\oplus)$. We obtain the number density of stars, $n_\mathrm{star}(D_d)$, along any given line of sight from the galaxy model described in \citet{Zhu2017}.

We will also make the simplification that $D_s \to \infty$, i.e. the sources are infinitely far away. We note that, at least up to an order of magnitude, our result for the optical depth does not strongly depend on the distance $D_s$, once $D_s$ is larger than about $15$\,kpc, i.e. past the galactic bulge. Since most of the number density of stars is in the galactic bulge, any pulsar that might be lensed by FFPs will have $D_s > 15\,$kpc. Thus, taking $D_s \to \infty$ is a reasonable approximation for both FRBs, which are extragalctic in origin, and pulsars. With this, we compute the integral in Eq.~\ref{eq:odepth} for the line of sight towards the galactic centre ($(l, b) = (0,0)$ in galactic coordinates) to be
\begin{equation}
    \tau \sim 10^{-9}.
\end{equation}
So, at any given time, a coherent source of radio emission located towards the galactic centre has a one in $10^9$ chance of being significantly lensed by a free-floating planet in the galaxy. Note that since we have taken all FFPs to have the same mass of $1\,M_\oplus$, the mass does not have to be integrated over, and the optical depth simply varies linearly with the assumed mass of the FFPs. 

\subsubsection{FFP event rate for pulsars}
\label{sec:pulsar_rate}

We now want to estimate the rate at which FFPs will significantly lens a given type of source. In this section we will calculate the event rate for pulsars. Pulsars and FFPs in the galaxy will generically be in motion relative to each other. Typical values for the relative angular speed for objects within the galaxy will be on the order of 10\,mas/year \citep{gaudi_microlensing_2012}. Given the typical masses of FFPs, the typical value of the parameter $s$ for pulsars lensed by FFPs will be $\sim 1$, and so $\theta_F \sim \theta_E$. In particular, the two time scales, $t_E$ and $t_F$, will both be on the order of 0.1\,days, which sets the time scale for the duration of a microlensing event in which a pulsar is lensed by an FFP. 

In the previous section we calculated the optical depth of FFPs towards the galactic centre, since most of the lenses will be located in the galactic centre. For a pulsar to have a significant probability of being lensed by an FFP, it must be located on the other side of the galactic centre from Earth. The majority of known pulsars are located within 8\,kpc of Earth, i.e. they are on the wrong side of galaxy for being lensed by FFPs in the galactic centre. However, a large fraction of all pulsars are located on the other side of the galaxy in a small 1\,degree by 40\,degrees sliver on the sky around the galactic centre \citep{2015aska.confE..40K}. Many future telescopes, such as HIRAX, will be able to observe and monitor these more distant pulsars. Since these pulsars make up a large fraction of the total pulsars in the galaxy, the optical depth computed in the previous section is, to an order of magnitude, the probability that a given pulsar is being significantly lensed by an FFP, at any given time. 

To get an event rate from this, we assume again that the maximum angular separation of source and lens to give rise to a detectable event in the wave-optics regime is $y_*^2 \theta_E$. Thus the probability that a source will undergo a lensing event in a time $dt$ is just the solid angle of the rectangle with width $2 y_*^2 \theta_E$ and length $\mu_\mathrm{rel.} dt$. This solid angle is the differential cross section in time, and we can rewrite it as $2y_*^2 \theta_E \mu_\mathrm{rel.} dt = 2 y_*^2 \theta^2_E dt / t_E = 2 \sigma_{\rm wave} dt / \pi y_*^2 t_E$, where $\sigma_{\rm wave}$ is given by Eq.~\ref{eq:sigwave-large}. Now to find the rate we can simply replace the cross section in Eq.~\ref{eq:odepth} by this differential cross section, and divide by $dt$. Thus, we can compute the rate for any given pulsar at a distance $D_s$ to undergo a microlensing event due to an FFP to be
\begin{equation}
    \Gamma(D_s) = \frac{2}{\pi y_*^2 t_E} \frac{\int_\mathcal{S} \tau(D_s; l, b) d\Omega}{\int_\mathcal{S} d\Omega},
    \label{eq:rate}
\end{equation}
where $\tau$ is the optical depth computed from Eq.~\ref{eq:odepth}, and $\mathcal{S}$ is the region of the sky where the sources are localized. For pulsars, we take $\mathcal{S}$ to be the 1\,degree by 40\,degrees region centred on the galactic centre. Evaluating Eq.~\ref{eq:rate} for pulsars, assuming again for simplicity that the pulsars are all located at infinity, we find $\Gamma \sim 10^{-7}$ per pulsar per day. The total event rate will be roughly this $\Gamma$ multiplied by the number of pulsars along lines-of-sight in $\mathcal{S}$. There are $\sim 10^3$ known pulsars, which leads to a total event rate of $\Gamma_\mathrm{tot.} \sim 10^{-4}\,\mathrm{day}^{-1}$, or $\Gamma_\mathrm{tot.} \sim 0.01\,\mathrm{year}^{-1}$.

Note, however, that the plasma propagation of the pulsar signal can lead to hundreds of effective images \citep[see e.g.][]{Brisken2010}, which na{\"i}vely would boost the event rate by two orders of magnitudes. However, if the flux is distirbuted evenly between the images, each image carries only 1\,percent of the total flux of the pulsar. If the flux is distributed evenly among $N$ images, $n$ of which are magnified by a factor $k$, then, in the geometric limit, the total flux is modulated by $nk/N$, so that there is no effective boost in cross section. However, since we observe an interference pattern of all the images, the amplitude of the flux is modulated by roughly $n^2 k/N$. Thus, for pulsars with hundreds of effective images, instead of a two-orders-of-magnitude boost to the cross section, the effects of plasma propagation yield a more modest factor of 10 increase. Taking this into account gives an estimate of the total event rate of $\Gamma_\mathrm{tot.} \sim 0.1\,\mathrm{year}^{-1}$.

This event rate is, of course, for a hypothetical survey that is always monitoring pulsars toward the galactic centre. Experiments such as the Canadian Hydrogen Intensity Mapping Experiment (CHIME) whose field of view is the entire northern sky will never see the galactic centre, and will thus have a much lower event rate for FFPs lensing pulsars. The Hydrogen Intensity and Real-time Analysis eXperiment \citep[HIRAX;][]{HIRAX}, which is under construction in the southern hemisphere, will see the galactic centre a significant fraction of the time, and so could potentially have an event rate approaching $\Gamma_\mathrm{tot.} \sim 0.1\,\mathrm{year}^{-1}$. Similarly, an envisioned future interferometer such as the Packed Ultra-Wideband Mapping Array \citep{Ansari:2018ury,PUMA}, which could monitor all pulsars discovered by SKA, has the potential to detect such events. Currently running lower frequency surverys, such as the Murchison Widefield Array \citep{2013PASA...30...31B}, which has a frequency range of 70-300\,MHz, also have the potential to observe these events, but for different mass ranges. For $\nu \sim 100\,$MHz, the requirement that $0.06 \lesssim s \lesssim 1$ gives a mass range of $[1, 1000]\,M_\oplus$. Thus, lower frequency observations can potentially probe higher mass FFP populations.

\subsubsection{FFP event rate for FRBs}
\label{sec:FRB_rate}

Since the duration of an FRB is of order milliseconds, which is much less than the order days time scale for microlensing events by FFPs in the galaxy, FRBs and FFPs are effectively static with respect to each other. As a result, the event rate will simply be the optical depth of FFPs toward the galactic centre, multiplied by the rate of FRB detections in that direction. CHIME has an FRB detection rate of about ten per day, across the whole sky; however, since the galactic centre never falls within CHIME's field of view, the event rate for CHIME would be vanishingly small. HIRAX will have a comparable FRB detection rate to CHIME, and will see the galactic centre. For HIRAX, the FRB detection rate towards the 1 degree by 40 degrees region centred on the galactic centre will be the total detection rate multiplied by the fraction of the sky taken up by this sliver. Multiplying this number by the optical depth for FFPs, $\tau \sim 10^{-9}$, gives an event rate of one event every $10^9$\,years. Even for next generation experiments such as PUMA, which anticipates an FRB detection rate of $10^3-10^4$ per day, the event rate will still only be one event per $10^6-10^7$\,years. However, we note that many future surveys will have adjustable exposure times, for example, using the DSA-2000 radio survey camera \citep{2019BAAS...51g.255H}, allowing for increased sensitivity. If these surveys choose to monitor the plane, the event rate may increase significantly. 

\subsection{Microlensing by stars in the Milky Way}
\label{sec:stars}

In Section~\ref{sec:planets}, we considered lenses that would produce microlensing events of radio sources for which both the geometric and Eikonal limits fail. For a single point-mass lens, this corresponds to $s \lesssim 1$, which results in a mass range that matches the masses of free-floating planets. However, even as we increase $s$ above 1, the Eikonal limit becomes a valid approximation, and wave-optical effects remain important, so long as the diffraction peaks are resolvable (see, e.g., Fig.~\ref{fig:schrwz_lcs}). For solar mass objects and radio sources with $\mu_{\rm rel.} \sim 10\,$mas/year, the scale of the separation of the diffraction peaks in time is on the order of $10\,$seconds. Thus, for pulsars, microlensing events will potentially exhibit an observable interference pattern for masses well above the mass range of FFPs. In the Eikonal limit, lenses are point-like if they are small with respect to the Einstein radius. For a solar-mass object 10\,kpc away, the Einstein radius is of order 1\,mas. Stars at this distance with radius of order $1\,R_\odot$ have an angular size of order $1\,\mu$as. Thus, stars will act as point lenses in the Eikonal limit for background pulsars. 

Since the cross section for a pulsar lensed by a point lens scales like $\sigma \sim \theta_E^2 \sim M$, the cross section for a solar mass object will be $10^6$ times larger than an earth mass object. Since in the previous section, we computed the optical depth of FFPs by assuming that FFPs were as numerous as stars in the galaxy, the result for the optical depth for stars is the same as for FFPs, but will be boosted by this factor of $10^6$ due to the larger cross section. Thus, the optical depth for stars is $\tau \sim 10^{-3}$. On the other hand, $t_E \sim M^{1/2}$, and so is a factor of $10^3$ larger for a solar mass object than an earth mass object. Using this to translate to a rate in the same way as we did in Section~\ref{sec:pulsar_rate} for FFPs, we find that the rate for stars to lens pulsars in the Eikonal limit is roughly $\Gamma \sim 10\,$per pulsar per year. Since in the Eikonal limit the timescale for a lensing event is given by $t_E$, and since for stars $t_E \sim 0.1\,$years, this suggests that roughly at any given time a pulsar located on the opposite side of the galactic centre is being magnified by $\gtrsim 10\%$ via lensing by a star in the galaxy. To detect such an effect, distant pulsars located towards the galactic centre would need to be monitored periodically for durations on the order of months. Experiments such as HIRAX and PUMA, which propose to regularly monitor pulsars, may indeed have the potential to detect the lensing of pulsars by stars in the Milky Way. 

For solar mass objects, the time delay between images will be of order $10\,\mu$s. This leads to two potential observational issues. Firstly, from Eq.~\ref{eq:wosc} the period of the oscillations in the frequency domain of the dynamic spectrum will be $\Delta \omega_{\rm osc} \sim 10\,$kHz. For an experiment such as CHIME which has frequency channels with bandwidth $\Delta \omega \sim 100\,$kHz \citep{Amiri:2018qsq}, the oscillations will be averaged over in the frequency domain. Secondly, in order to observe an interference pattern in the time domain, the time delay must be smaller than the coherence time, which for each frequency channel is given by the inverse of the bandwidth, $t_c \equiv 2\pi \Delta \omega^{-1} \sim 100\,\mu$s. Since the time delay increases linearly with the lens mass, for larger solar mass objects the interference pattern in the time domain will be washed out. Both of these issues can be resolved by increasing the frequency resolution to be less than $\sim 10\,$kHz. This can be achieved by taking the Fourier transform along the time domain for discrete bins of width $\Delta t$ that is greater than the inverse of the desired frequency resolution; in this case, $\Delta t \gtrsim 10^{-4}\,$s. The CHIME FRB backend buffers raw voltage data with a $2.56\,\mu$s cadence \citep{Amiri:2018qsq}, so that a sub-$10\,$kHz frequency resolution over the appropriate frequency ranges can easily be achieved. This procedure naturally reduces the time resolution; however, since the scale of separation of the diffraction peaks in time is $\sim 10\,$s, the time resolution can be decreased significantly without losing the ability to resolve these peaks. In practice, for pulsar scintillation observations, the interference pattern is not generally measured in the dynamic spectrum, but rather in the secondary spectrum. This is the case for pulsar scintillation due to scattering in the ISM; despite the large, order-millisecond time delays, scintles are easily observable in the secondary spectrum \citep[see e.g.][]{Brisken2010}.

\subsection{Extra-galactic point lenses}
\label{sec:exgal_lenses}

One may also be interested in extra-galactic point lenses. Such lenses would, naturally, only lens extra-galactic sources, and so we must abandon pulsars as our paradigmatic background source. FRBs, however, are perfect candidates for extra-galactic sources of coherent emission, since they are located at distances of $D_s \sim \,\,$Gpc \citep{2019Sci...365..565B, dongzi2019}. However, we must be careful when applying the formulae in Section~\ref{sec:point_lens} to determine how an FRB will be lensed by an intervening point mass, since FRBs manifest as short, irregularly repeating bursts, when they repeat at all. Eq.~\ref{eq:F} is the diffraction integral for a single monochromatic plane wave, and thus, effectively, assumes that the source has been shining forever. For a burst of finite size, this assumption only holds when the time delay is small relative to the burst duration. In the case of point lenses, the time delay is on the order of
\begin{equation}
    4GM = 1.97 \times 10^{-2} \, \left(\frac{M}{M_\odot} \right)\,\mathrm{ms}.
\end{equation}
Since FRBs are order-millisecond bursts, point masses less than a solar mass will satisfy the condition that the time delay be much shorter than the FRB. So, point masses in the range $[0.1\,M_\oplus, 1\,M_\odot]$ will significantly lens FRBs, with magnifications given by Eq.~\ref{eq:Hwave}. The lower bound on this mass range is obtained from the same, distance independent argument used to find the mass range in Section~\ref{sec:planets}. Planets fall within this mass range, justifying our previous use of Eq,~\ref{eq:Hwave} and the associated cross section for FRBs lensed by FFPs in Section~\ref{sec:planets}. 

For lenses larger than this mass range, the time delay begins to exceed the duration of the FRB. In this case, instead of having a single burst with an intensity modulation determined by Eq.~\ref{eq:Hwave}, the FRB will manifest as a single burst followed by an echo, with time delay dependent on the mass of the lens and the separation of the source and the lens. The brightness ratio between the original image and the echo is dependent only on the separation (Eq.~\ref{eq:pnt_12}), and so a measurement of the brightness ratio and the time delay will yield the mass of the lens. \citet{munoz_lensing_2016} has proposed using microlensing of FRBs in this regime to constrain the amount of dark matter contained in massive compact halo objects (MACHOs) in the mass range of $[20, 100]\,M_\odot$. 

For objects within the mass range $[0.1\,M_\oplus, 1\,M_\odot]$, the intensity modulation of an FRB will be determined by Eq.~\ref{eq:Hwave}, which we have shown also provides a means of measuring the mass of the lens. To determine when the interference pattern will be observable, we estimate the Fresnel scale for extra-galactic lenses and FRBs is $\theta_F \sim 10^{-5}\,$mas, by taking the lens distance to be $D \sim 1\,{\rm Gpc}$. Assuming that for extra-galactic lenses, $\mu_\mathrm{rel.}D_d \sim 10^3\,$km/s, a typical value for the peculiar velocity of nearby galaxies \citep{2014MNRAS.445.2677S}, then $\mu_\mathrm{rel.} \sim 10^{-7}$mas/year, and so $t_F \sim 10^{2}$\,years. Thus, the duration of the Fresnel crossing time is much longer than the duration of an FRB, so that over the duration of a single burst FRBs will be effectively static with respect to extra-galactic lenses as in Section~\ref{sec:FRB_rate}

It is of interest to compute the microlensing event rate of FRBs by extra-galactic point lenses. For a general population of lenses falling within the mass range $[0.1\,M_\oplus, 1\,M_\odot]$, we can compute the total optical depth on the sky from \citep{Schneider}
\begin{equation}
    \tau(z_s) = \frac{c}{H_0} \frac{1}{D^2_s} \int_0^{z_s} dz \int dM \hat{\sigma}(D_d, D_s, M) n(D_d, M) D^2_d \frac{(1+z)^{-2}}{E(z)},
\end{equation}
where $\hat{\sigma} = \sigma_{\rm wave} D^2_d$, and $n(D_d, M)$ is the number density of point-like lenses with mass $M$ at a distance $D_d$. Since FRBs are at cosmological distances, we perform the integral over redshift, so that $D_d = D_A(z)$ and $D_s = D_A(z_s)$, where $D_A$ is the angular diamater distance as a function of redshift. Note that because the distances are now angular diamater distances, the distance $D_{ds}$, which appears in $\theta_E$, is no longer simply $D_s - D_d$. The term $(1+z)^{-2}/E(z)$ is $dR/dz$, where $R$ is the proper distance, and $E(z) = \sqrt{\Omega_m (1+z)^3 + \Omega_r (1+z)^4 + \Omega_k (1+z)^2 + \Omega_\Lambda}$. 

For a non-evolving population, the density is just
\begin{equation}
    n(D_d, M) = (1+z)^3 n_0(M),
\end{equation}
where $n_0(M)$ is the number density of point-like lenses with mass $M$ today. The density of a non-evolving population as a function of redshift is then obtained by holding the density constant within a comoving volume. 

As an example of such a calculation, we can compute the optical depth for extra-galactic FFPs. In order to compute the optical depth, we again make the assumption that the FFPs all have mass $1\,M_\oplus$, since we do not know the mass distribution of FFPs in our galaxy, let alone other galaxies. We estimate $n_0$ to be the local number density of galaxies multiplied by the average number of FFPs per galaxy, $n_0 = n_\mathrm{gal.} N_{FFP}$. The local number density of galaxies is roughly $n_\mathrm{gal.} \sim 0.1\,\mathrm{Mpc}^{-3}$ \citep{2016ApJ...830...83C}, and we take $N_{FFP} \sim 10^{11}$, which is the number of FFPs in the Milky Way, assuming FFPs are as numerous as stars. With the simplifying assumption that all FRBs are located at $D_s \sim 1\,$Gpc, or $z_s \approx 0.3$, and again taking $A_H = 10^{-1}$, we find that for extra-galactic FFPs the optical depth is $\tau \sim 10^{-7}$. For a CHIME detection rate of ten FRBs per day, this translates to an event rate of roughly $10^{-4}$\,per year. Thus, current experiments do not have a high enough FRB event rate for the possibility of using microlensing to detect extra-galactic exoplanets. However, for next generation experiments such as PUMA, which may have an FRB detection rate of as much as two orders of magnitude higher than CHIME, a detection of such an event may enter the realm of possibility. 

Recently, \citet{Katz2019} argue that diffractive effects in the lensing of FRBs will also be able to provide strong constraints on dark matter in the form of MACHOs in the $10^{-4}$ to $0.1\,M_\odot$ mass range. With new FRB surveys on the horizon, observations of the gravitational lensing of FRBs with diffractive effects have the potential to be powerful probes of a variety of objects at cosmological distances, for a large mass range. 

\section{Conclusion}
\label{sec:concl}

In the coming years, a wealth of data from new observations of pulsars and FRBs will be taken by current and next-generation radio telescopes. Pulsars and FRBs will generically be lensed by intervening gravitational potentials, and being coherent sources of radio emission, are likely to exhibit wave effects in their lensing. In this paper, we have shown that not only do wave effects break degeneracies that would otherwise be present in the geometric limit of optics, they result in a significant boost to the cross section of lensing events. While we demonstrated this boost in cross section for the specific case of a point lens, it was simply a result of constructive interference between images increasing the magnification further from the lens, which will be a generic feature of many different lenses. We also showed that galactic point masses in the range of $0.1$ to $100\,M_\oplus$ will lens radio sources in the full wave-optics regime, beyond both the geometric and Eikonal limit. Isolated objects in this mass range will include FFPs. We computed the rate for pulsars to be lensed by FFPs in the galaxy and found that it could reach as high as $\Gamma \sim 0.1\,$year, suggesting that pulsars could be a useful probe of the population of FFPs in the galaxy. Similarly, FRBs, which are located at cosmological distances, will exhibit interference effects in their lensing for a wide range of masses, from $10^{-6}\,M_\odot$ to $1\,M_\odot$, which includes FFPs. We have argued that the lensing of an FRB, which is a short, transient phenomenon, is sufficient to measure the mass of the lens, since the mass can be inferred from interference effects at a single time slice. 

Microlensing studies often assume the geometric limit of optics. However, it is important to check when wave effects are important. The geometric limit fails near caustic crossings, and for coherent low frequency sources of light. This failure is in fact a benefit, as diffractive effects can provide additional physical information about the lens, and strongly enhance the detectability of a microlensing event. Accounting for these effects, including in situations beyond the point-mass limit studied here, could very well open up new frontiers of discovery in the transient radio sky.

\section*{Data Availability}
No new data were generated or analysed in support of this research.

\section*{Acknowledgements}

We thank Ramesh Bhat, Katelyn Breivik, Neal Dalal, Job Feldbrugge, and Scott Gaudi for useful discussions. Research at the Perimeter Institute is supported in part by the Government of Canada through the Department of Innovation, Science and Economic Development Canada and by the Province of Ontario through the Ministry of Economic Development, Job Creation and Trade. We receive support from the Ontario Research Fund - Research Excellence
Program (ORF-RE), the Canadian Institute for Advanced Research (CIFAR),
the Canadian Foundation for Innovation (CFI), the Simons Foundation, Thoth.
Technology Inc, and Alexander von Humboldt Foundation. We acknowledge
the support of the Natural Sciences and Engineering Research Council of
Canada (NSERC), [funding reference number 523638-201, RGPIN-2019-067,
523638-201]. Cette recherche a \'{e}t\'{e} financ\'{e}e par le Conseil de recherches
en sciences naturelles et en g\'{e}nie du Canada (CRSNG), [num\'{e}ro de
r\'{e}f\'{e}rence 523638-201, RGPIN-2019-067, 523638-201].




\bibliographystyle{mnras_sjf}
\bibliography{biblio} 

\begin{thebibliography}{}
\makeatletter
\relax
\def\mn@urlcharsother{\let\do\@makeother \do\$\do\&\do\#\do\^\do\_\do\%\do\~}
\def\mn@doi{\begingroup\mn@urlcharsother \@ifnextchar [ {\mn@doi@}
  {\mn@doi@[]}}
\def\mn@doi@[#1]#2{\def\@tempa{#1}\ifx\@tempa\@empty \href
  {http://dx.doi.org/#2} {doi:#2}\else \href {http://dx.doi.org/#2} {#1}\fi
  \endgroup}
\def\mn@eprint#1#2{\mn@eprint@#1:#2::\@nil}
\def\mn@eprint@arXiv#1{\href {http://arxiv.org/abs/#1} {{\tt arXiv:#1}}}
\def\mn@eprint@dblp#1{\href {http://dblp.uni-trier.de/rec/bibtex/#1.xml}
  {dblp:#1}}
\def\mn@eprint@#1:#2:#3:#4\@nil{\def\@tempa {#1}\def\@tempb {#2}\def\@tempc
  {#3}\ifx \@tempc \@empty \let \@tempc \@tempb \let \@tempb \@tempa \fi \ifx
  \@tempb \@empty \def\@tempb {arXiv}\fi \@ifundefined
  {mn@eprint@\@tempb}{\@tempb:\@tempc}{\expandafter \expandafter \csname
  mn@eprint@\@tempb\endcsname \expandafter{\@tempc}}}

\bibitem[\protect\citeauthoryear{Amiri et~al.}{Amiri
  et~al.}{2018}]{Amiri:2018qsq}
Amiri M.,  et~al., 2018,  \mn@eprint {arXiv} {1803.11235}

\bibitem[\protect\citeauthoryear{Ansari et~al.}{Ansari
  et~al.}{2018}]{Ansari:2018ury}
Ansari R.,  et~al., 2018,  \mn@eprint {arXiv} {1810.09572}

\bibitem[\protect\citeauthoryear{{Bannister} et~al.,}{{Bannister}
  et~al.}{2019}]{2019Sci...365..565B}
{Bannister} K.~W.,  et~al., 2019, \mn@doi [Science] {10.1126/science.aaw5903},
  \href {https://ui.adsabs.harvard.edu/abs/2019Sci...365..565B} {365, 565},
  \mn@eprint {arXiv} {1906.11476}

\bibitem[\protect\citeauthoryear{Barnacka, Glicenstein  \& Moderski}{Barnacka
  et~al.}{2012}]{Barnacka:2012bm}
Barnacka A.,  Glicenstein J.~F.,   Moderski R.,  2012, \mn@doi [Phys. Rev.]
  {10.1103/PhysRevD.86.043001}, D86, 043001,  \mn@eprint {arXiv} {1204.2056}

\bibitem[\protect\citeauthoryear{{Bowman} et~al.,}{{Bowman}
  et~al.}{2013}]{2013PASA...30...31B}
{Bowman} J.~D.,  et~al., 2013, \mn@doi [PASA] {10.1017/pas.2013.009}, \href
  {https://ui.adsabs.harvard.edu/abs/2013PASA...30...31B} {30, e031},
  \mn@eprint {arXiv} {1212.5151}

\bibitem[\protect\citeauthoryear{{Brisken}, {Macquart}, {Gao}, {Rickett},
  {Coles}, {Deller}, {Tingay}  \& {West}}{{Brisken} et~al.}{2010}]{Brisken2010}
{Brisken} W.~F.,  {Macquart} J.~P.,  {Gao} J.~J.,  {Rickett} B.~J.,  {Coles}
  W.~A.,  {Deller} A.~T.,  {Tingay} S.~J.,   {West} C.~J.,  2010, \mn@doi [APJ]
  {10.1088/0004-637X/708/1/232}, \href
  {https://ui.adsabs.harvard.edu/abs/2010ApJ...708..232B} {708, 232},
  \mn@eprint {arXiv} {0910.5654}

\bibitem[\protect\citeauthoryear{{Conselice}, {Wilkinson}, {Duncan}  \&
  {Mortlock}}{{Conselice} et~al.}{2016}]{2016ApJ...830...83C}
{Conselice} C.~J.,  {Wilkinson} A.,  {Duncan} K.,   {Mortlock} A.,  2016,
  \mn@doi [APJ] {10.3847/0004-637X/830/2/83}, \href
  {https://ui.adsabs.harvard.edu/abs/2016ApJ...830...83C} {830, 83},
  \mn@eprint {arXiv} {1607.03909}

\bibitem[\protect\citeauthoryear{Dai, Li, Zackay, Mao  \& Lu}{Dai
  et~al.}{2018}]{Dai:2018enj}
Dai L.,  Li S.-S.,  Zackay B.,  Mao S.,   Lu Y.,  2018, \mn@doi [Phys. Rev.]
  {10.1103/PhysRevD.98.104029}, D98, 104029,  \mn@eprint {arXiv} {1810.00003}

\bibitem[\protect\citeauthoryear{Deguchi \& Watson}{Deguchi \&
  Watson}{1986a}]{PhysRevD.34.1708}
Deguchi S.,  Watson W.~D.,  1986a, \mn@doi [Phys. Rev. D]
  {10.1103/PhysRevD.34.1708}, 34, 1708

\bibitem[\protect\citeauthoryear{{Deguchi} \& {Watson}}{{Deguchi} \&
  {Watson}}{1986b}]{1986ApJ...307...30D}
{Deguchi} S.,  {Watson} W.~D.,  1986b, \mn@doi [\apj] {10.1086/164389}, \href
  {https://ui.adsabs.harvard.edu/abs/1986ApJ...307...30D} {307, 30}

\bibitem[\protect\citeauthoryear{{Diego}}{{Diego}}{2019}]{Diego2019}
{Diego} J.~M.,  2019, arXiv e-prints, \href
  {https://ui.adsabs.harvard.edu/abs/2019arXiv191105736D} {p.
  arXiv:1911.05736},  \mn@eprint {arXiv} {1911.05736}

\bibitem[\protect\citeauthoryear{{Diego}, {Hannuksela}, {Kelly}, {Pagano},
  {Broadhurst}, {Kim}, {Li}  \& {Smoot}}{{Diego}
  et~al.}{2019}]{DiegoHannuksela2019}
{Diego} J.~M.,  {Hannuksela} O.~A.,  {Kelly} P.~L.,  {Pagano} G.,  {Broadhurst}
  T.,  {Kim} K.,  {Li} T.~G.~F.,   {Smoot} G.~F.,  2019, \mn@doi [AAP]
  {10.1051/0004-6361/201935490}, \href
  {https://ui.adsabs.harvard.edu/abs/2019A&A...627A.130D} {627, A130},
  \mn@eprint {arXiv} {1903.04513}

\bibitem[\protect\citeauthoryear{{Feldbrugge}, {Pen}  \& {Turok}}{{Feldbrugge}
  et~al.}{2019}]{job_pl}
{Feldbrugge} J.,  {Pen} U.-L.,   {Turok} N.,  2019, arXiv e-prints, \href
  {https://ui.adsabs.harvard.edu/abs/2019arXiv190904632F} {p.
  arXiv:1909.04632},  \mn@eprint {arXiv} {1909.04632}

\bibitem[\protect\citeauthoryear{{Foreman-Mackey}, {Hogg}, {Lang}  \&
  {Goodman}}{{Foreman-Mackey} et~al.}{2013}]{emcee}
{Foreman-Mackey} D.,  {Hogg} D.~W.,  {Lang} D.,   {Goodman} J.,  2013, \mn@doi
  [PASP] {10.1086/670067}, \href
  {https://ui.adsabs.harvard.edu/abs/2013PASP..125..306F} {125, 306},
  \mn@eprint {arXiv} {1202.3665}

\bibitem[\protect\citeauthoryear{Gaudi}{Gaudi}{2012}]{gaudi_microlensing_2012}
Gaudi B.~S.,  2012, \mn@doi [Annu. Rev. Astron. Astrophys.]
  {10.1146/annurev-astro-081811-125518}, 50, 411

\bibitem[\protect\citeauthoryear{{Gould}}{{Gould}}{1992}]{Gould:1992}
{Gould} A.,  1992, \mn@doi [Astrophys. J. Lett.] {10.1086/186279}, 386, L5

\bibitem[\protect\citeauthoryear{{Gould} \& {Loeb}}{{Gould} \&
  {Loeb}}{1992}]{1992ApJ...396..104G}
{Gould} A.,  {Loeb} A.,  1992, \mn@doi [\apj] {10.1086/171700}, \href
  {https://ui.adsabs.harvard.edu/abs/1992ApJ...396..104G} {396, 104}

\bibitem[\protect\citeauthoryear{{Hallinan} et~al.,}{{Hallinan}
  et~al.}{2019}]{2019BAAS...51g.255H}
{Hallinan} G.,  et~al., 2019, in BAAS. p.~255 \mn@eprint {arXiv} {1907.07648}

\bibitem[\protect\citeauthoryear{Heyl}{Heyl}{2010a}]{Heyl:2010dg}
Heyl J.~S.,  2010a,  \mn@eprint {arXiv} {1002.3007}

\bibitem[\protect\citeauthoryear{Heyl}{Heyl}{2010b}]{Heyl:2009av}
Heyl J.,  2010b, \mn@doi [Mon. Not. Roy. Astron. Soc.]
  {10.1111/j.1745-3933.2009.00795.x}, 402, 39,  \mn@eprint {arXiv} {0910.3922}

\bibitem[\protect\citeauthoryear{Heyl}{Heyl}{2011}]{Heyl:2010hm}
Heyl J.~S.,  2011, \mn@doi [Mon. Not. Roy. Astron. Soc.]
  {10.1111/j.1365-2966.2010.17814.x}, 411, 1787,  \mn@eprint {arXiv}
  {1003.0250}

\bibitem[\protect\citeauthoryear{Jaroszynski \& Paczynski}{Jaroszynski \&
  Paczynski}{1995}]{Jaroszynski:1995cd}
Jaroszynski M.,  Paczynski B.,  1995, \mn@doi [Astrophys. J.] {10.1086/176593},
  455, 443,  \mn@eprint {arXiv} {astro-ph/9503043}

\bibitem[\protect\citeauthoryear{{Kaspi} \& {Kramer}}{{Kaspi} \&
  {Kramer}}{2016}]{2016arXiv160207738K}
{Kaspi} V.~M.,  {Kramer} M.,  2016, arXiv e-prints, \href
  {https://ui.adsabs.harvard.edu/abs/2016arXiv160207738K} {p.
  arXiv:1602.07738},  \mn@eprint {arXiv} {1602.07738}

\bibitem[\protect\citeauthoryear{Katz, Kopp, Sibiryakov  \& Xue}{Katz
  et~al.}{2018}]{Katz:2018zrn}
Katz A.,  Kopp J.,  Sibiryakov S.,   Xue W.,  2018, \mn@doi [JCAP]
  {10.1088/1475-7516/2018/12/005}, 1812, 005,  \mn@eprint {arXiv} {1807.11495}

\bibitem[\protect\citeauthoryear{{Katz}, {Kopp}, {Sibiryakov}  \& {Xue}}{{Katz}
  et~al.}{2019}]{Katz2019}
{Katz} A.,  {Kopp} J.,  {Sibiryakov} S.,   {Xue} W.,  2019, arXiv e-prints,
  \href {https://ui.adsabs.harvard.edu/abs/2019arXiv191207620K} {p.
  arXiv:1912.07620},  \mn@eprint {arXiv} {1912.07620}

\bibitem[\protect\citeauthoryear{{Keane} et~al.,}{{Keane}
  et~al.}{2015a}]{2015aska.confE..40K}
{Keane} E.,  et~al., 2015a, in Advancing Astrophysics with the Square Kilometre
  Array (AASKA14). p.~40 \mn@eprint {arXiv} {1501.00056}

\bibitem[\protect\citeauthoryear{Keane et~al.}{Keane
  et~al.}{2015b}]{Keane:2014vja}
Keane E.~F.,  et~al., 2015b, \mn@doi [PoS] {10.22323/1.215.0040}, AASKA14, 040,
   \mn@eprint {arXiv} {1501.00056}

\bibitem[\protect\citeauthoryear{{Li}, {Yalinewich}  \& {Breysse}}{{Li}
  et~al.}{2019}]{dongzi2019}
{Li} D.,  {Yalinewich} A.,   {Breysse} P.~C.,  2019, arXiv e-prints, \href
  {https://ui.adsabs.harvard.edu/abs/2019arXiv190210120L} {p.
  arXiv:1902.10120},  \mn@eprint {arXiv} {1902.10120}

\bibitem[\protect\citeauthoryear{{Ma}, {Mao}, {Ida}, {Zhu}  \& {Lin}}{{Ma}
  et~al.}{2016}]{Ma2016}
{Ma} S.,  {Mao} S.,  {Ida} S.,  {Zhu} W.,   {Lin} D. N.~C.,  2016, \mn@doi
  [MNRAS] {10.1093/mnrasl/slw110}, \href
  {https://ui.adsabs.harvard.edu/abs/2016MNRAS.461L.107M} {461, L107},
  \mn@eprint {arXiv} {1605.08556}

\bibitem[\protect\citeauthoryear{Macquart et~al.}{Macquart
  et~al.}{2015}]{Macquart:2015uea}
Macquart J.~P.,  et~al., 2015.  \mn@eprint {arXiv} {1501.07535}

\bibitem[\protect\citeauthoryear{{Mao} \& {Paczynski}}{{Mao} \&
  {Paczynski}}{1991}]{1991ApJ...374L..37M}
{Mao} S.,  {Paczynski} B.,  1991, \mn@doi [\apjl] {10.1086/186066}, \href
  {https://ui.adsabs.harvard.edu/abs/1991ApJ...374L..37M} {374, L37}

\bibitem[\protect\citeauthoryear{Montero-Camacho, Fang, Vasquez, Silva  \&
  Hirata}{Montero-Camacho et~al.}{2019}]{Montero-Camacho:2019jte}
Montero-Camacho P.,  Fang X.,  Vasquez G.,  Silva M.,   Hirata C.~M.,  2019,
  \mn@doi [JCAP] {10.1088/1475-7516/2019/08/031}, 1908, 031,  \mn@eprint
  {arXiv} {1906.05950}

\bibitem[\protect\citeauthoryear{{Mr{\'o}z} et~al.,}{{Mr{\'o}z}
  et~al.}{2017}]{Mroz2017}
{Mr{\'o}z} P.,  et~al., 2017, \mn@doi [NAT] {10.1038/nature23276}, \href
  {https://ui.adsabs.harvard.edu/abs/2017Natur.548..183M} {548, 183},
  \mn@eprint {arXiv} {1707.07634}

\bibitem[\protect\citeauthoryear{{Mr{\'o}z} et~al.,}{{Mr{\'o}z}
  et~al.}{2018}]{2018AJ....155..121M}
{Mr{\'o}z} P.,  et~al., 2018, \mn@doi [\aj] {10.3847/1538-3881/aaaae9}, \href
  {https://ui.adsabs.harvard.edu/abs/2018AJ....155..121M} {155, 121},
  \mn@eprint {arXiv} {1712.01042}

\bibitem[\protect\citeauthoryear{Mu{\~n}oz, Kovetz, Dai  \&
  Kamionkowski}{Mu{\~n}oz et~al.}{2016}]{munoz_lensing_2016}
Mu{\~n}oz J.~B.,  Kovetz E.~D.,  Dai L.,   Kamionkowski M.,  2016, \mn@doi
  [Phys. Rev. Lett.] {10.1103/PhysRevLett.117.091301}, 117, 091301

\bibitem[\protect\citeauthoryear{{Nakamura} \& {Deguchi}}{{Nakamura} \&
  {Deguchi}}{1999}]{nakamura}
{Nakamura} T.~T.,  {Deguchi} S.,  1999, \mn@doi [Progress of Theoretical
  Physics Supplement] {10.1143/PTPS.133.137}, \href
  {https://ui.adsabs.harvard.edu/abs/1999PThPS.133..137N} {133, 137}

\bibitem[\protect\citeauthoryear{Newburgh et~al.}{Newburgh
  et~al.}{2016}]{HIRAX}
Newburgh L.~B.,  et~al., 2016, \mn@doi [Proc. SPIE Int. Soc. Opt. Eng.]
  {10.1117/12.2234286}, 9906, 99065X,  \mn@eprint {arXiv} {1607.02059}

\bibitem[\protect\citeauthoryear{Ng}{Ng}{2017}]{Ng:2017djg}
Ng C.,  2017, \mn@doi [IAU Symp.] {10.1017/S1743921317010638}, 337, 179,
  \mn@eprint {arXiv} {1711.02104}

\bibitem[\protect\citeauthoryear{Niikura et~al.}{Niikura
  et~al.}{2019}]{Niikura:2019zjd}
Niikura H.,  et~al., 2019, \mn@doi [Nat. Astron.] {10.1038/s41550-019-0723-1},
  3, 524,  \mn@eprint {arXiv} {1701.02151}

\bibitem[\protect\citeauthoryear{{Paczynski}}{{Paczynski}}{1986}]{Paczynski1986}
{Paczynski} B.,  1986, \mn@doi [APJ] {10.1086/164140}, \href
  {https://ui.adsabs.harvard.edu/abs/1986ApJ...304....1P} {304, 1}

\bibitem[\protect\citeauthoryear{{Petroff}, {Hessels}  \& {Lorimer}}{{Petroff}
  et~al.}{2019}]{Petroff2019}
{Petroff} E.,  {Hessels} J.~W.~T.,   {Lorimer} D.~R.,  2019, \mn@doi [AAPR]
  {10.1007/s00159-019-0116-6}, \href
  {https://ui.adsabs.harvard.edu/abs/2019A&ARv..27....4P} {27, 4},  \mn@eprint
  {arXiv} {1904.07947}

\bibitem[\protect\citeauthoryear{{Schneider}, {Ehlers}  \& {Falco}}{{Schneider}
  et~al.}{1992}]{Schneider}
{Schneider} P.,  {Ehlers} J.,   {Falco} E.~E.,  1992, {Gravitational Lenses},
  \mn@doi{10.1007/978-3-662-03758-4.
}

\bibitem[\protect\citeauthoryear{{Slosar} et~al.,}{{Slosar}
  et~al.}{2019}]{PUMA}
{Slosar} A.,  et~al., 2019, in BAAS. p.~53 \mn@eprint {arXiv} {1907.12559}

\bibitem[\protect\citeauthoryear{{Springob} et~al.,}{{Springob}
  et~al.}{2014}]{2014MNRAS.445.2677S}
{Springob} C.~M.,  et~al., 2014, \mn@doi [MNRAS] {10.1093/mnras/stu1743}, \href
  {https://ui.adsabs.harvard.edu/abs/2014MNRAS.445.2677S} {445, 2677},
  \mn@eprint {arXiv} {1409.6161}

\bibitem[\protect\citeauthoryear{Sugiyama, Kurita  \& Takada}{Sugiyama
  et~al.}{2019}]{Sugiyama:2019dgt}
Sugiyama S.,  Kurita T.,   Takada M.,  2019,  \mn@eprint {arXiv} {1905.06066}

\bibitem[\protect\citeauthoryear{{Sumi} et~al.,}{{Sumi}
  et~al.}{2011}]{Sumi2011}
{Sumi} T.,  et~al., 2011, \mn@doi [NAT] {10.1038/nature10092}, \href
  {https://ui.adsabs.harvard.edu/abs/2011Natur.473..349S} {473, 349},
  \mn@eprint {arXiv} {1105.3544}

\bibitem[\protect\citeauthoryear{Takahashi \& Nakamura}{Takahashi \&
  Nakamura}{2003}]{Takahashi:2003ix}
Takahashi R.,  Nakamura T.,  2003, \mn@doi [Astrophys. J.] {10.1086/377430},
  595, 1039,  \mn@eprint {arXiv} {astro-ph/0305055}

\bibitem[\protect\citeauthoryear{Ulmer \& Goodman}{Ulmer \&
  Goodman}{1995}]{Ulmer:1994ij}
Ulmer A.,  Goodman J.,  1995, \mn@doi [Astrophys. J.] {10.1086/175422}, 442,
  67,  \mn@eprint {arXiv} {astro-ph/9406042}

\bibitem[\protect\citeauthoryear{{Vanderlinde} et~al.,}{{Vanderlinde}
  et~al.}{2019}]{2019arXiv191101777V}
{Vanderlinde} K.,  et~al., 2019, arXiv e-prints, \href
  {https://ui.adsabs.harvard.edu/abs/2019arXiv191101777V} {p.
  arXiv:1911.01777},  \mn@eprint {arXiv} {1911.01777}

\bibitem[\protect\citeauthoryear{Wyrzykowski et~al.,}{Wyrzykowski
  et~al.}{2011}]{Wyrzykowski2011}
Wyrzykowski L.,  et~al., 2011, \mn@doi [MNRAS]
  {10.1111/j.1365-2966.2010.18150.x}, 413, 493,  \mn@eprint {}
  {http://oup.prod.sis.lan/mnras/article-pdf/413/1/493/18589162/mnras0413-0493.pdf}

\bibitem[\protect\citeauthoryear{{Zhu} et~al.,}{{Zhu} et~al.}{2017}]{Zhu2017}
{Zhu} W.,  et~al., 2017, \mn@doi [AJ] {10.3847/1538-3881/aa8ef1}, \href
  {https://ui.adsabs.harvard.edu/abs/2017AJ....154..210Z} {154, 210},
  \mn@eprint {arXiv} {1701.05191}

\bibitem[\protect\citeauthoryear{{de Forcrand}}{{de
  Forcrand}}{2010}]{2010arXiv1005.0539D}
{de Forcrand} P.,  2010, arXiv e-prints, \href
  {https://ui.adsabs.harvard.edu/abs/2010arXiv1005.0539D} {p. arXiv:1005.0539},
   \mn@eprint {arXiv} {1005.0539}

\makeatother
\end{thebibliography}



\appendix


\bsp	
\label{lastpage}
\end{document}